\pdfoutput=1
\documentclass[12pt,a4paper]{article}
\usepackage{amsmath,amsfonts,amssymb,amsthm,bbm,bm}
\usepackage{pdfsync}
\usepackage{graphicx}
\usepackage[latin1]{inputenc}
\usepackage{hyperref}
\usepackage[all]{xy}
\usepackage{stmaryrd}
\usepackage{rotating}
\usepackage{color}  
\usepackage{slashed}
\usepackage[dvipsnames]{xcolor}
\usepackage{comment}
\usepackage{cite}
\newcommand{\bc}{\begin{center}}
\newcommand{\ec}{\end{center}}

\newcommand{\cL}{{\mathcal L}}

\newcommand{\cO}{{\mathcal O}}

\newcommand{\be}{\begin{align}}
\newcommand{\ee}{\end{align}}
\newcommand{\bea}{\begin{eqnarray}}
\newcommand{\eea}{\end{eqnarray}}
\newcommand{\bs}{\begin{subequations}}
\newcommand{\es}{\end{subequations}}
\newcommand{\nn}{\nonumber}

\newcommand{\tl}{\tilde}
\def\p{\partial}

\newcommand{\dd}{\mathrm{d}}

\newcommand{\pd}{\partial}

\newcommand{\Mijrad}[1]{\!\!\stackrel{(#1)}{M}\!{}^{\text{rad}}_{\!ij}}

\def\eps{\epsilon}

\def\th{\theta}

\def\m{\mu}
\def\n{\nu}
\def\mn{{\mu\nu}}

\allowdisplaybreaks

\title{{\bf Multipole Expansion of Gravitational Waves: \\from Harmonic to Bondi coordinates}\\[0.15cm] \large (or ``Monsieur de Donder meets Sir Bondi'')
}

\date{}

\begin{document}
%%%%%%%%%%%%%%%%%%%%%%%%%%%%%%%%%%%%%%%%

\maketitle

\vspace{-1.5cm}

\centerline{\large{\bf{Luc Blanchet,$^{a}$\footnote{luc.blanchet@iap.fr} Geoffrey Comp\`{e}re,$^{b}$\footnote{gcompere@ulb.ac.be}}}}\vspace{6pt}

\centerline{\large{\bf{Guillaume Faye,$^{a}$\footnote{faye@iap.fr} Roberto Oliveri,$^{c}$\footnote{roliveri@fzu.cz} Ali Seraj$^{\,b}$\footnote{aseraj@ulb.ac.be}}}}

\bigskip\medskip
\centerline{\textit{{}$^{a}$ $\mathcal{G}\mathbb{R}\varepsilon{\mathbb{C}}\mathcal{O}$, Institut d'Astrophysique de Paris, UMR 7095,}}
\centerline{\textit{CNRS \& Sorbonne Universit{\'e}, 98\textsuperscript{bis} boulevard Arago, 75014 Paris, France}}

\medskip
\centerline{\textit{{}$^{b}$ Universit\'{e} Libre de Bruxelles, Centre for Gravitational Waves, }}
\centerline{\textit{International Solvay Institutes, CP 231, B-1050 Brussels, Belgium}}

\medskip
\centerline{\textit{{}$^{c}$ CEICO, Institute of Physics of the Czech Academy of Sciences,}}
\centerline{\textit{Na Slovance 2, 182 21 Praha 8, Czech Republic}}
\vspace{1cm}

%----------------------------------------------------------------------------
\begin{abstract}
\noindent 
We transform the metric of an isolated matter source in the multipolar post-Minkowskian approximation from harmonic (de Donder) coordinates to radiative Newman-Unti (NU) coordinates. To linearized order, we obtain the NU metric as a functional of the mass and current multipole moments of the source, valid all-over the exterior region of the source. Imposing appropriate boundary conditions we recover the generalized Bondi-van der Burg-Metzner-Sachs residual symmetry group. To quadratic order, in the case of the mass-quadrupole interaction, we determine the contributions of gravitational-wave tails in the NU metric, and prove that the expansion of the metric in terms of the radius is regular to all orders. The mass and angular momentum aspects, as well as the Bondi shear, are read off from the metric. They are given by the radiative quadrupole moment including the tail terms.
\end{abstract}

%----------------------------------------------------------------------------

\setcounter{footnote}{0}% Reset footnote counter

\newpage
\tableofcontents

\section{Introduction}
\label{sec:intro}

\subsection{Motivations}

Gravitational waves (GWs), whose physical existence was controversial for years, were established rigorously in the seminal works of Bondi, van der Burg, Metzner and Sachs~\cite{1962RSPSA.269...21B, 1962RSPSA.270..103S}. The Bondi-Sachs formalism describes the asymptotic structure near future null infinity of the field generated by isolated self-gravitating sources. This asymptotic structure was further elucidated thanks to the tools of the Newman-Penrose formalism~\cite{Newman:1961qr} and conformal compactifications~\cite{Penrose63} leading to the concept of asymptotically simple spacetimes in the sense of Penrose~\cite{Penrose65}. Asymptotically simple spacetimes are now proven to follow from large sets of initial data which are stationary at spatial infinity, see \textit{e.g.} the review~\cite{Friedrich:2017cjg}.  

Bondi coordinates or Bondi tetrad frames are defined from an outgoing light cone congruence with radial sections parametrized by the luminosity (areal) distance. A variant of these coordinates are the Newman-Unti (NU) coordinates whose radial coordinate is instead an affine parameter~\cite{newman1963class}. Bondi gauge and NU gauge share all essential features and can easily be mapped to each other~\cite{Barnich:2011ty,Barnich:2013ba}. Under the assumption of asymptotic simplicity, Einstein's equations admit a consistent asymptotic solution~\cite{Tamburino:1966zz, Barnich:2010eb}. Such an asymptotic series is however limited to the vicinity of null infinity and it does not, in particular, resolve the source that generates the radiation. 

Recent interest in Bondi gauge arose from the fact that it is preserved under an infinite set of residual symmetries, dubbed the generalized BMS group, that is generated by supertranslations and arbitrary diffeomorphisms on the two-sphere~\cite{deBoer:2003vf, Barnich:2009se, Barnich:2010eb, Barnich:2011mi, Campiglia:2014yka, Campiglia:2015yka}, which gives rise to two infinite sets of flux-balance laws \cite{Barnich:2010eb, Barnich:2011ty, Barnich:2011mi, Barnich:2013axa, Strominger:2013jfa, Strominger:2014pwa, Flanagan:2015pxa, Compere:2016hzt, Hawking:2016sgy, Barnich:2016lyg, Compere:2018ylh,Nichols:2017rqr, Nichols:2018qac, Bonga:2018gzr, Distler:2018rwu, Barnich:2019vzx, Ashtekar:2019viz, Ashtekar:2019rpv, Compere:2019gft}. Thanks to junction conditions at spatial infinity~\cite{Strominger:2013jfa}, the generalized BMS group is a symmetry of the quantum gravity S-matrix, which gives rise to Ward identities that are identical to Weinberg' soft graviton theorem~\cite{He:2014laa} and to the subleading soft graviton theorem~\cite{Kapec:2014opa, Campiglia:2014yka}. 

For GW generation and applications to the data analysis of the GW events one needs the connection between the asymptotic structure of the field and explicit matter sources. This is achieved by the multipolar post-Minkowskian (MPM) expansion~\cite{Blanchet:1985sp, Blanchet:1986dk, Blanchet:1987wq, Blanchet:1992br} which combines the multipole expansion for the field in the exterior region of the source with a nonlinearity expansion in powers of the gravitational constant $G$. The MPM formalism is defined in harmonic coordinates, also known as de Donder coordinates. At linear order the MPM expansion reduces to the linear metric as written by Thorne~\cite{Thorne:1980ru} and is characterized in terms of two infinite sets of \textit{canonical} multipole moments, namely the mass and current multipole moments. A class of radiative coordinate systems exists such that the MPM expansion leads to asymptotically simple spacetimes for sources that are stationary before some given time in the remote past~\cite{Blanchet:1986dk}. In such radiative coordinates, two infinite sets of \textit{radiative} multipoles can be defined in terms of the canonical multipoles. They parametrize the asymptotic transverse-traceless waveform or, equivalently, the two polarizations of the Bondi shear. 

In addition, the MPM formalism has to be matched to the post-Newtonian (PN) field in the near-zone and the interior of the source, which allows us to express the canonical multipoles in terms of the actual \textit{source} multipoles and, furthermore, yields the radiation-reaction forces caused by the radiation onto their sources~\cite{Blanchet:1998in, PB02, BFN05}. The MPM-PN formalism was applied to compact binary systems and permitted to compute the GW phase evolution of inspiralling compact binaries to high PN order, see notably~\cite{BFIJ02, BDEI04, Blanchet:2008je, Faye:2014fra}.

The main objective of this paper is to make explicit the relationship between Bondi expansions and the MPM formalism. The Bondi and NU gauges belong to the general class of radiative gauges in the sense of~\cite{Madore, Blanchet:1986dk}. Here we will describe the construction of the explicit diffeomorphism transforming the metric in the MPM expansion from harmonic coordinates to NU coordinates. The diffeomorphism is perturbative in powers of $G$ and, for each PM order, it is valid everywhere outside the source. After imposing standard boundary conditions, we find this diffeomorphism to be unique up to generalized BMS transformations~\cite{deBoer:2003vf, Barnich:2009se,  Barnich:2010eb, Barnich:2011mi, Campiglia:2014yka, Campiglia:2015yka, Campiglia:2016hvg, Compere:2018ylh, Campiglia:2020qvc, Compere:2020lrt}, as we will cross-check in details. This allows us to transpose known results on the exterior MPM metric in harmonic gauge for a particular multipolar mode coupling and a given post-Minkowskian order to a metric in NU gauge, written as an exact expression to all orders in the radial expansion. As an illustration, we will explicitly derive the Bondi metric of the second-order post-Minkowskian (2PM or $G^2$) perturbation corresponding to mass-quadrupole interactions~\cite{Blanchet:1992br, Blanchet:1997jj}. In particular this entails the description of GW tails within the Bondi asymptotic framework.

The rest of the paper is organized as follows. Section~\ref{sec:not} is devoted to our notation and conventions. Section~\ref{sec:linear} recalls the harmonic-coordinates description of the metric in terms of canonical moments at linearized order. In Sec.~\ref{sec:algo} we present an algorithm implementing the transformation from harmonic coordinates to NU coordinates. In Sec.~\ref{sec:NUlinear} we derive the NU metric as a function of the mass and current multipoles to linearized order. In Sec.~\ref{sec:BMS} we impose standard boundary conditions for the asymptotic metric and naturally recover from our algorithm the gauge freedom associated with the BMS group. Notably, in Sec.~\ref{sec:aspects}, we obtain the Bondi mass aspect, the angular momentum aspect and the Bondi shear as multipole expansions parametrized by the canonical moments. In Sec.~\ref{sec:quadratic} we apply the algorithm to the quadratic metric (\textit{i.e.} to 2PM order in the MPM formalism), focusing on the quadratic interaction between the mass monopole and the mass quadrupole. Explicit results on GW tails obtained in harmonic coordinates, are then conveyed into the NU metric in Sec.~\ref{sec:tail}, to any order in the radial expansion. Finally we discuss in Sec.~\ref{sec:loss} the mass and angular momentum GW losses in the Bondi-NU framework at the level of the quadrupole-quadrupole interaction. The paper ends with a short conclusion and perspectives in Sec.~\ref{sec:concl}. Two appendices gather technical details on the map between Bondi and NU gauges (\ref{app:map}), and the all-order PM formul\ae\, for the coordinate change equations (\ref{app:PM}).

\subsection{Notation and conventions}\label{sec:not}

We adopt units with the speed of light $c=1$. The Newton gravitational constant $G$ is kept explicit to bookmark post-Minkowskian (PM) orders. We will refer to lower case Latin indices from $a$ to $h$ as indices on the two-dimensional sphere, while lower case Latin indices from  $i$ to $z$ will refer to three-dimensional Cartesian indices. The Minkowski metric is $\eta_{\mu\nu}=\text{diag}(-1,+1,+1,+1)$.

We denote Cartesian coordinates as $x^{\mu} = (t, \mathbf{x})$ and spherical ones as $(t, r, \theta^a)$. Here, the radial coordinate $r$ is defined as $r=\vert\mathbf{x}\vert$ and $\theta^a=(\theta, \varphi)$ with  $a,b,\dots =\{1,2\}$. The unit directional vector is denoted as $n^i = n^i(\theta^a) = x^i/r$. Euclidean spatial indices $i, j, \dots = \{1,2,3\}$ are raised and lowered with the Kronecker metric $\delta_{ij}$. Furthermore, we define the Minkowskian outgoing vector $k^{\mu} \partial_{\mu} = \partial_{t} + n^i \partial_i$ with $\partial_i=\partial/\partial x^i$, or, in components, $k^{\mu} = (1, n^i)$ and $k_{\mu} = (-1, n^i)$. In retarded spherical coordinates $(u,r,\theta^a)$ with $u=t-r$, we have $k^\mu \partial_\mu = \partial_r\big\vert_{u}$. We employ the natural basis on the unit 2-sphere $e_a=\frac{\pd}{\pd \theta^a}$ embedded in $\mathbb R^3$  with components $e^i_a = \partial n^i/\partial\th^a$. Given  the unit metric on the sphere $\gamma_{ab} = \mathrm{diag}(1,\sin^2\th)$ we have: $n^i e^i_a = 0$, $\partial_i \theta^a = r^{-1} \gamma^{ab} e_b^i$, $\gamma_{ab} = \delta_{ij} e^i_a e^j_b$ and $\gamma^{ab} e^i_a e^j_b  = \perp^{ij}$, where $\perp^{ij}=\delta^{ij}-n^i n^j$ is the projector onto the sphere. We also use the notation $e_{\langle a}^i e_{b\rangle}^j=e^i_{( a}e^j_{b)}-\frac{1}{2}\gamma_{ab}\!\perp^{ij}$ for the trace-free product of basis vectors. Introducing the covariant derivative $D_a$ compatible with the sphere metric, $D_a \gamma_{bc}=0$, we have $D_a e^i_b = D_b e^i_a = D_a D_b n^i = - \gamma_{ab} n^i$. 

Given a general manifold, harmonic/de Donder coordinates are specified by using a tilde: $\tl x^{\mu} = (\tl t, \mathbf{\tl x})$ or $(\tl t, \tl r, \tl \theta^a)$. The metric tensor is $\tl g_{\mu\nu}(\tl x)$. Asymptotically flat spacetimes admit as a background structure the Minkowskian outgoing vector $\tl k^{\mu} = (1, \tl n^i)$, the basis on the sphere $\tl e^i_a = \partial \tl n^i/\partial\tl \th^a$, \textit{etc}. We define the retarded time $\tl u$ in harmonic coordinates as $\tl u=\tl t- \tl r$, such that $\tl k_\mu = - \tl \partial_{\mu} \tl u$.

Newman-Unti (NU) coordinates are denoted $x^{\mu} = (u, r, \theta^a)$ with $\theta^a=(\theta, \varphi)$. The metric tensor in NU coordinates is denoted as $g_{\mu\nu}(x)$, with all other notation, such as the natural basis on the sphere $e^i_a$ and the metric $\gamma_{ab}$, as previously.

We denote by $L=i_1i_2 \dots i_\ell$ a multi-index made of $\ell$ spatial indices. We use short-hands for: the multi-derivative operator $\partial_L = \partial_{i_1}\dots\partial_{i_\ell}$, the product of vectors $n_L=n_{i_1}\dots n_{i_\ell}$ and $x_L=x_{i_1}\dots x_{i_\ell}=r^\ell n_{L}$. The multipole moments $M_L$ and $S_L$ are symmetric and trace-free (STF). The transverse-trace-free (TT) projection operator is denoted $\perp^{ijkl}_\mathrm{TT} = \perp^{k(i}\perp^{j)l} - \frac{1}{2}\!\perp^{ij}\perp^{kl}$. Time derivatives are indicated by superscripts $(q)$ or by dots.

%----------------------------------------------------------------------------
\section{From harmonic gauge to Newman-Unti gauge}
\label{sec:harmNU}
%----------------------------------------------------------------------------

\subsection{Linear metric in harmonic coordinates}
\label{sec:linear}

We work with the gothic metric deviation defined as $h^\mn=\sqrt{\vert \tl g\vert}\tl g^\mn-\eta^\mn$ and satisfying the de Donder (or harmonic) gauge condition $\tl \pd_\m h^\mn=0$. The Einstein field equations in harmonic coordinates read as
\begin{align}\label{eq:EFE}
\tl \Box h^\mn = \frac{16\pi G}{c^4}\vert \tl g\vert T^\mn + \Lambda^\mn(h,\partial h,\partial^2h)\,,
\end{align}
where $\tl\Box=\tl\Box_\eta$ is the flat d'Alembertian operator, and the right-hand side contains the matter stress-energy tensor $T^\mn$ as well as the back-reaction from the metric itself, in the form of an infinite sum $\Lambda^\mn$ of quadratic or higher powers of $h$ and its space-time derivatives. We shall consider the metric generated by an isolated matter system, in the form of a non-linearity or post-Minkowskian (PM) expansion, labeled by $G$,
\begin{align}\label{eq:PM}
h^\mn=\sum_{n=1}^{+\infty}G^n h_{n}^\mn\,.
\end{align}
Furthermore we consider the metric in the \textit{vacuum} region outside the isolated matter system, and assume that each PM coefficient $h_{n}^\mn$ in Eq.~\eqref{eq:PM} is in the form of a multipole expansion, parametrized by so-called canonical multipole moments. We call this the multipolar-post-Minkowskian approximation~\cite{Blanchet:1985sp}. In the linearized approximation the vacuum Einstein field equations in harmonic coordinates read $\tl\Box h^\mn_{1}=\tl\partial_\nu h^\mn_{1}=0$, whose most general retarded solution, modulo an infinitesimal harmonic gauge transformation, is~\cite{Thorne:1980ru} 
\begin{subequations}\label{eq:linearmetric}
\begin{align}
    h_{1}^{00} &= -4 \sum_{\ell=0}^{+\infty}\frac{(-)^{\ell}}{\ell!}\tl\p_L\left(\frac{M_{L}(\tl u)}{\tl r} \right)\,, \\
    h_{1}^{0j} &= 4\sum_{\ell=1}^{+\infty} \frac{(-)^{\ell}}{\ell!}\left[\tl\p_{L-1}\left(\frac{M^{(1)}_{j L-1}(\tl u)}{\tl r} \right) + \frac{\ell}{\ell+1}\tl\p_{p L-1}\left(\frac{\varepsilon_{jpq}S_{q L-1}(\tl u)}{\tl r} \right)\right]\,, \\
    h_{1}^{jk} &=  -4\sum_{\ell=2}^{+\infty} \frac{(-)^{\ell}}{\ell!}\left[\tl\p_{L-2}\left(\frac{M^{(2)}_{jk L-2}(\tl u)}{\tl r} \right) +  \frac{2\ell}{\ell+1}\tl\p_{p L-2}\left(\frac{\varepsilon_{pq(j}S^{(1)}_{k)q L-2}(\tl u)}{\tl r} \right)\right]\,,
\end{align}
\end{subequations}
given in terms of symmetric-trace-free (STF) canonical mass and current multipole moments $M_L$ and $S_L$ depending on the harmonic coordinate retarded time $\tl u = \tl t - \tl r$. Among these moments, the mass monopole $M$ is the constant (ADM) total mass of the system, $P^i=M^{(1)}_i$ is the constant linear momentum and $S_i$ is the constant angular momentum. We can expand the linear metric in powers of $1/\tl r$ using the formula (valid for arbitrary STF tensors $M_L$)
\begin{subequations}\label{eq:formula}
\begin{align}
\tl\p_L\left(\frac{M_{L}(\tl u)}{\tl r} \right)&=(-)^\ell\,\tl n_L\sum_{k=0}^{\ell}a_{k\ell} \,\frac{M_L^{(\ell-k)}(\tl u)}{\tl r^{k+1}}\,,\\
\text{with}\quad a_{k\ell} &= \frac{(\ell+k)!}{2^k k!(\ell-k)!}\,.
\end{align}
\end{subequations}

A method has been proposed in~\cite{Blanchet:1985sp} to compute each of the PM coefficients up to any order $n$, starting from the linear metric~\eqref{eq:linearmetric}. Each of the PM approximation is then obtained as a functional of the canonical multipole moments $M_L$ and $S_L$. The construction represents the most general solution of the Einstein field equations outside a matter source without any incoming flux from past null infinity. This is the so-called MPM formalism. The relation between the canonical moments and the source moments depending on actual source parameters is known~\cite{Blanchet:1998in, PB02, BFN05}. 

In this paper we assume that the metric is stationary in the past in the sense that all the multipole moments are constant before some finite instant in the past, say $M_L(\tl u)=\text{const}$ and $S_L(\tl u)=\text{const}$ when $\tl u \leqslant -\mathcal{T}$. Under this assumption all non-local integrals we shall meet  will be convergent at their bound in the infinite past.\footnote{This assumption may be weakened to the situation where the source is initially made of free particles moving on unbound hyperbolic like orbits (initial scattering). In this case we would have $M_L(\tl u) \sim (-\tl u)^\ell$ and $ S_L(\tl u) \sim (-\tl u)^\ell$ when $\tl u\to-\infty$, and the tail integrals in the radiative moment, Eq.~\eqref{eq:radtailL} below, would still be convergent for such initial state~\cite{BS93}.}

%----------------------------------------------------------------------------
\subsection{Algorithm to transform harmonic to NU metrics}\label{sec:algo}
%---------------------------------------------------------------------------

Consistently with the PM expansion~\eqref{eq:PM}, we assume that the NU coordinates are related to the harmonic coordinates by the following class of transformations
\begin{subequations}\label{eq:transfBondi}
\begin{align}
    u &= \tl u +\sum_{n=1}^{+\infty}G^{n} U_{n}(\tl u,\tl r, \tl \theta^a)\,,\\[-0.15cm]
    r &= \tl r + \sum_{n=1}^{+\infty}G^{n} R_{n}(\tl u,\tl r, \tl \theta^a)\,,\\[-0.15cm]
    \theta^a &= \tl \theta^a + \sum_{n=1}^{+\infty}G^{n} \Theta^a_{n}(\tl u,\tl r, \tl \theta^b)\,,
\end{align}
\end{subequations}
where the PM coefficients $U_{n}$, $R_{n}$ and $\Theta^a_{n}$ are functions of the harmonic coordinates $(\tl u,\tl r, \tl \theta^a)$ to be determined, with $\tl u = \tl t - \tl r$.

The NU gauge\footnote{The NU and Bondi gauges differ by a choice of the radial coordinate. See more details in~\cite{Barnich:2011ty} and in the Appendix~\ref{app:map} below.} is defined by the following conditions:
\begin{equation}\label{eq:NU gauge0}
g_{ur}=-1\,, \qquad  g_{rr} = 0\,, \qquad  g_{ra} = 0\,.
\end{equation}
For computational reasons, it is more convenient to work with the inverse metric components, for which the NU gauge reads as
\begin{equation}\label{eq:NU gaugeinv}
g^{uu}=0\,, \qquad g^{ur} = -1\,, \qquad g^{ua} = 0\,,
\end{equation}
The gauge is constructed such that (i) the outgoing vector $k_\mu=-\pd_\mu u$ is null, (ii) the angular coordinates are constant along null rays $k^\mu\pd_\mu \theta^a=0$, and (iii) the radial coordinate is an affine parameter on outgoing null curves, \emph{i.e.} $k^\mu\pd_\mu r =1$. The strategy to construct the perturbative diffeomorphism is the following. From the NU gauge conditions~\eqref{eq:NU gaugeinv}, one finds the following constraints on the transformation laws~\eqref{eq:transfBondi}, namely
\begin{subequations}\label{eq:NU gauge}
\begin{align}
 \tl g^\mn(\tl x) \frac{\p u}{\p \tl x^{\mu}} \,\frac{\p u}{\p \tl x^{\nu}} &= 0\,,\\  
 \tl g^\mn(\tl x) \frac{\p u}{\p \tl x^{\mu}}\,\frac{\p r}{\p \tl x^{\nu}} &= -1\,, \\  
 \tl g^\mn(\tl x) \frac{\p u}{\p \tl x^{\mu}}\,\frac{\p \theta^a}{\p \tl x^{\nu}} &= 0\,.
\end{align}\end{subequations}
Inserting the linear metric~\eqref{eq:linearmetric} this permits to solve for the linear corrections $U_{1}$, $R_{1}$ and $\Theta^a_{1}$, modulo an arbitrariness related \emph{in fine} to BMS transformations. Then one uses
\begin{subequations}
\begin{align}
g^{rr}(x) &= \tl g^\mn(\tl x) \frac{\p r}{\p \tl x^{\mu}} \,\frac{\p r}{\p \tl x^{\nu}} \,,\\  
g^{ra}(x) &= \tl g^\mn(\tl x) \frac{\p r}{\p \tl x^{\mu}} \,\frac{\p \theta^a}{\p \tl x^{\nu}} \,,\\  
g^{ab}(x) &= \tl g^\mn(\tl x) \frac{\p \theta^a}{\p \tl x^{\mu}} \,\frac{\p \theta^b}{\p \tl x^{\nu}} \,,
\end{align}\end{subequations}
to deduce $g_{uu}=-g^{rr}+g^{ra}g_{ua}$, $g_{ua}=g^{rb}g_{ab}$, $g_{ab}=(g^{ab})^{-1}$ to linear order. We can then read off, respectively, the Bondi mass aspect, the Bondi angular momentum aspect and the Bondi shear. To quadratic order one inserts the metric $h_2^{\mu\nu}(\tl x)$ in harmonic coordinates solving Eq.~\eqref{eq:EFE} to order $G^2$, and obtain $U_{2}$, $R_{2}$, $\Theta^a_{2}$ and the NU metric to order $G^2$. In the end we have to re-express the metric in terms of NU coordinates using the inverse of Eq.~\eqref{eq:transfBondi}. This algorithm can be iterated in principle at any arbitrary order in powers of $G$.

\section{Newman-Unti metric to linear order}
%\label{sec:linear}

\subsection{Solving the NU gauge conditions}\label{sec:NUlinear}

At linear order in $G$, the constraints~\eqref{eq:NU gauge} are equivalent to the following equations for the linear coefficients $U_{1}$, $R_{1}$ and $\Theta^a_{1}$, involving the directional derivative along the direction $\tl k^{\mu} = (1, \tl n^i)$ of the Minkowski null cone:
\begin{subequations}\label{linearized functions}
\begin{align}
\tl k^{\mu}\tl\partial_\mu U_{1} &= \frac{1}{2} \tl k_\mu \tl k_\nu h^{\mu\nu}_{1}\,,\\
\tl k^{\mu}\tl\partial_\mu R_{1} &= -\frac{1}{2} \tl n_i \tl n_j h^{ij}_{1} + \frac{1}{2} h^{ii}_{1} - \dot{U}_{1}\,,\\
\tl k^{\mu}\tl\partial_{\mu}\Theta_{1}^{a} &= \frac{\tl e_{i}^a}{\tl r}\Bigl(\tl\partial_i U_{1} - \tl k_\mu h_{1}^{\mu i} \Bigr)\,.
\end{align}
\end{subequations}
where the overdot denotes the derivative with respect to $\tilde{u}$. Notice that $h^{ii}_{1}=0$ for the metric~\eqref{eq:linearmetric}. Using the explicit form of the linearized metric~\eqref{eq:linearmetric}--\eqref{eq:formula} one readily obtains the most general solutions of those equations as 
\begin{subequations}\label{eq:U1R1Th1}
\begin{align}
U_{1} &= -2\bigl(M - \tl n_iP_i\bigr)\ln (\tl r/\mathcal{P}) + 4\sum_{\ell=1}^{+\infty}\frac{1}{\ell!}\sum_{k=1}^{\ell}\frac{(2k-1)a_{k\ell}}{(\ell+k-1)(\ell+k)} \frac{\tl n_L M_L^{(\ell-k)}}{\tl r^k} -  \xi_1^{u}\,,\label{eq:U1}\\
R_{1} &= M + \left[3-2\ln (\tl r/\mathcal{P})\right] \tl n_i P_i + 2\sum_{\ell=2}^{+\infty}\frac{1}{\ell!}\sum_{k=1}^{\ell-1}\frac{(\ell-k)(\ell+3k+1)a_{k\ell}}{(\ell+k)(\ell+k-1)(k+1)} \frac{\tl n_L M_L^{(\ell-k)}}{\tl r^k} -  \xi_1^{r}\,,\label{eq:R1}\\
\Theta_{1}^a &= \frac{\tl e^a_i}{\tl r}\biggl[ 2P_i \bigl[1-\ln (\tl r/\mathcal{P}) \bigr] \\
& - 4\sum_{\ell=1}^{+\infty}\frac{1}{\ell!}\sum_{k=1}^{\ell}\frac{a_{kl}}{(\ell+k)(k+1)}\frac{\tl n_{L-1}}{\tl r^{k}}\left(\frac{2k^2-\ell}{\ell+k-1} M_{iL-1}^{(\ell-k)} + \frac{2 k \ell}{\ell+1} \varepsilon_{ijk} \tl n_j S_{kL-1}^{(\ell-k)} \right) \biggr] -  \xi^a_1\,,\nn
\end{align}
\end{subequations}
where $\mathcal{P}$ is an irrelevant constant. We recognize the standard logarithmic deviation $ u = \tl u -2G M \ln (\tl r/\mathcal{P}) + \mathcal{O}(\tl r^{-1})$ from harmonic to radiative coordinates; see \textit{e.g.}~\cite{Blanchet:1986dk}.

Furthermore we have added the most general homogeneous solution of the differential equations \eqref{linearized functions} denoted by $\xi^\mu$. These are indeed the residual  gauge transformations preserving the NU gauge \eqref{eq:NU gauge}, at linearized order, \textit{i.e.} $ x^\mu \to  x^\mu + \xi^\mu$ with $\xi^\mu = O(G^1)$. The linear gauge transformation, $\xi=G \xi_1$ takes the form
\begin{equation}\label{eq:gauge}
\xi^{u}_1 = f\,,\qquad \xi^{r}_1 =  - \tl r\dot{f}+Q\,, \qquad  \xi^a_1 = Y^{a} -\frac{1}{\tl r} \tl D^a f\,,
\end{equation}
where $f$, $Q$ and $Y^{a}$ are arbitrary functions of $\tl u=\tl t-\tl r$ and the angles $\tl \theta^a$. Note that  for later convenience, we made explicit into the expression of $R_{1}$ given by~Eq.~\eqref{eq:R1} some constant monopolar and dipolar ($\ell=0,1$) contributions corresponding to a redefinition of the radial coordinate as $ \tl r\to \tl r+ G (M+3 \tl n_iP_i)$, thanks to the arbitrary function $Q$ in Eq.~\eqref{eq:gauge}.

The metric in NU gauge is immediately obtained at linear order in $G$ from the linear metric $h_1^\mn$ (and its trace $h_1=\eta_\mn h_1^\mn = - h_1^{00}$) as given by Eq.~\eqref{eq:linearmetric}  together with the linear coefficients $U_{1}$, $R_{1}$ and $\Theta^a_{1}$ as 
\begin{subequations}\label{eq:metric0}
\begin{align}
 g_{uu} &= -1 + 2G\Bigl( \dot{R}_1 +\dot{U}_1 + \frac{1}{4}h_1 \Bigr) + \mathcal{O}(G^2)\,,\\
 g_{ua} &= G\, r^2 \Bigl[-\dot{\Theta}_{1a} +  r^{-1}e_a^i h_1^{0i} +  r^{-2}D_a\bigl(R_1+U_1 \bigr)\Bigr] + \mathcal{O}(G^2)\,,\\
 g_{ab} &= r^2 \gamma_{ab} - G \, r^2\Bigl( 2  D_{(a} \Theta_{1b)} + 2  r^{-1} \gamma_{ab} R_1 +  e_a^i  e_b^j h_1^{ij} - \frac{1}{2} \gamma_{ab} \,h_1 \Bigr) + \mathcal{O}(G^2)\,.
\end{align}
\end{subequations}
Note that the final result for the metric has been written in terms of the NU coordinates $x^\mu$. As a result the spatial metric $ g_{ab}$ is given by a covariant tensorial expression on the sphere, involving the Lie derivative $\cL_{\Theta_1} \gamma_{ab} = 2 D_{(a} \Theta_{1b)}$. To this end, we have written the leading contribution in the spatial metric $ g_{ab}$ in terms of NU coordinates to linear order in $G$ as 
\begin{equation}
{\tl r^2}\tl \gamma_{ab} =   r^2 \Bigl[\gamma_{ab} - 2G\Bigl( r^{-1} R_{1} \gamma_{ab} + \Theta_1^c  \Gamma^e_{c(a} \gamma_{b)e}\Bigr) \Bigr] + \mathcal{O}(G^2)\,,
\end{equation} 
where $ \Gamma^a_{bc}$ denotes the Christoffel symbol on the sphere. At linear order in $G$, we can equivalently replace the harmonic coordinates by the NU ones, as the correction will be at $\cO(G^2)$. Plugging the results~\eqref{eq:U1R1Th1} into the metric~\eqref{eq:metric0}, we find 
\begin{subequations}\label{eq:metric}
\begin{align}
 g_{uu} &= -1 + 2G \sum_{\ell=0}^{+\infty}\frac{(\ell+1)(\ell+2)}{\ell!}\sum_{k=0}^{\ell}\frac{a_{k\ell}}{(k+1)(k+2)} \frac{n_L M_L^{(\ell-k)}}{r^{k+1}} + \delta_\xi  g_{uu}+ \mathcal{O}(G^2)\,,\label{eq:metricuu}\\
%%%%%%%%%%%%%%%%%%%%%%%%%%%%%%%%%%%%%%%%%%%%%%%%%%%%%%%%%%%%%%%%%
 g_{ua} &= G\,e_a^i\biggl\{ - \sum_{\ell=2}^{+\infty}\frac{\ell+2}{\ell!}n_{L-1}\Bigl[  M_{iL-1}^{(\ell)} - \frac{2\ell}{\ell+1}\varepsilon_{ipq}n_p S_{qL-1}^{(\ell)}\Bigr] \nn\\
& + 2\sum_{\ell=1}^{+\infty}\frac{\ell+2}{\ell!}n_{L-1}\sum_{k=1}^{\ell}\frac{a_{k\ell}}{k+2} \frac{1}{r^k} \Bigl[M_{iL-1}^{(\ell-k)}+\frac{2\ell}{\ell+1}\varepsilon_{ipq}n_p S_{qL-1}^{(\ell-k)}\Bigr]\biggr\} + \delta_\xi  g_{ua}+ \mathcal{O}(G^2)\,,\label{eq:metricua}\\
%%%%%%%%%%%%%%%%%%%%%%%%%%%%%%%%%%%%%%%%%%%%%%%%%%%%%%%%%%%%%
 g_{ab} &= r^2 \Biggl[ \gamma_{ab} + 4G\,e_{\langle a}^i e_{b\rangle}^j \sum_{\ell=2}^{+\infty}\frac{1}{\ell!}\frac{n_{L-2}}{r}\biggl\{ M_{ijL-2}^{(\ell)}-\frac{2\ell}{\ell+1}\varepsilon_{ipq}n_p S_{jqL-2}^{(\ell)} \nn\\
&\qquad\qquad 
+ \sum_{k=2}^{\ell}\frac{k-1}{k+1} \frac{a_{k\ell}}{r^k} \Bigl[M_{ijL-2}^{(\ell-k)}+\frac{2\ell}{\ell+1}\varepsilon_{ipq}n_p S_{jqL-2}^{(\ell-k)}\Bigr]\biggr\} \Biggr] + \delta_\xi  g_{ab}+ \mathcal{O}(G^2)\,,\label{eq:metricab}
\end{align}
\end{subequations}
where we have posed $e_{\langle a}^i e_{b\rangle}^j=e^i_{( a}e^j_{b)}-\frac{1}{2}\gamma_{ab}\!\perp^{ij}$. The last terms correspond to the freedom left in the metric, which is associated with the gauge vector~\eqref{eq:gauge}, and are given by
\begin{subequations}\label{eq:deltametric}
\begin{align}
\delta_\xi g_{uu} &= 2G\Bigl[-\dot{Q} - \dot{f} + r\ddot{f}\Bigr]+ \mathcal{O}(G^2)\,,\label{delta guu}\\
\delta_\xi  g_{ua} &= G\Bigl[-D_a\bigl(Q + f\bigr) + r^2 \dot Y_{a}\Bigr]+ \mathcal{O}(G^2)\,,\label{delta gua}\\
\delta_\xi  g_{ab} &= 2G\Bigl[r^2\bigl(-\gamma_{ab}\dot{f}+ D_{(a}Y_{b)}\bigr) + r \bigl(-D_a D_b f+\gamma_{ab}Q\bigr)\Bigr]+ \mathcal{O}(G^2)\,.\label{delta gab}
\end{align}
\end{subequations}

\subsection{Boundary conditions and the BMS group}\label{sec:BMS}

An asymptotic frame is defined from boundary Dirichlet gauge fixing conditions, which pick a specific foliation by constant $ u$ surfaces and a specific measure on the codimension 2 boundary. The boundary gauge fixing conditions when $ r \to \infty$ are
\begin{subequations}\label{eq:bndgauge}
	\begin{align}
	 g_{uu} &= \mathcal{O}( r^{0})\,,\label{BC:guu}\\
	 g_{ua} &= \mathcal{O}( r^0)\,, \label{BC:gua}\\
	\det g_{ab} &= r^4 \sin^2\!\theta + \mathcal{O}( r^2)\,,\label{BC:gab} 
	\end{align}
\end{subequations}
where the first term in Eq.~\eqref{BC:gab} is the determinant of the metric on the unit sphere metric. Notice that Eq.~\eqref{BC:gab} not only fixes the measure on the sphere, but also requires that the shear which appears at order $\mathcal{O}(r^3)$ is trace-free, see the discussion around Eq.~(2.5) of~\cite{Barnich:2011ty}. The boundary condition \eqref{BC:gab} only determines the leading order determinant, which is compatible with Newmann-Unti gauge.\footnote{In contrast, the Bondi gauge fixing condition $\partial_r(\text{det}g_{ab}/r^4)=0$ fixes the determinant at any $r$, except at leading order.} The metric~\eqref{eq:metric}--\eqref{eq:deltametric} does not yet respect the boundary conditions~\eqref{eq:bndgauge}. Thus one has to implement an infinitesimal transformation in order to achieve the gauge with appropriate asymptotic behavior.  

The first condition~\eqref{BC:guu} implies that $\ddot{f}=0$, hence $\dot{f}$ must only be a function of the angles $ \theta^a$. The second condition~\eqref{BC:gua} implies that $\dot{Y}^a=0$, \textit{i.e.} that $Y^a$ also is only a function of the angles. To impose the last condition~\eqref{BC:gab}, we note that the leading metric on the sphere $ \gamma_{ab}$ already satisfies the leading behavior of~\eqref{BC:gab}, \textit{i.e.}, its measure is that of a unit metric on the sphere. Therefore the leading term in Eq.~\eqref{eq:deltametric} must be trace-free, thus $\dot f=\frac{1}{2}D_aY^a$, which is consistent with $\ddot{f}=0$. Similarly the next-to-leading term $\mathcal{O}( r)$ in $ g_{ab}$ must also be trace-free, hence $Q=\frac{1}{2} \Delta f$ where $\Delta=D^a D_a$ is the Laplacian on the sphere. Summarizing all these, we have
\begin{align}
Q=\dfrac{1}{2}\Delta f \,,\qquad  f(u,\theta^a)=T(\th^a)+\dfrac{u}{2} D_aY^a\,,\qquad Y^a=Y^a(\th^b)\,.
\end{align}
The simplest choice that brings the metric~\eqref{eq:metric} into the form~\eqref{eq:bndgauge} is of course obtained by setting $f=Q=T=Y^a=0$. This choice is generally assumed in the perturbative approach to gravitational waves in harmonic coordinates, see \textit{e.g.}~\cite{Blanchet:2018yqa}. However, after fulfilling all the conditions, \textit{i.e.} the gauge conditions~\eqref{eq:NU gauge0} and the asymptotic boundary conditions~\eqref{eq:bndgauge}, we are still left with the infinitesimal coordinate transformations generated by the gauge vector field $ \xi_\text{BMS}^\mu \equiv \xi^\mu$, with components
\begin{subequations}
\begin{align}\label{BMSgen}
     \xi_\text{BMS}^u &= T + \dfrac{u}{2} D_aY^a\,,\\ 
     \xi_\text{BMS}^r &=-\dfrac{r}{2}D_a Y^a + \frac{1}{2} \Delta\left(T + \dfrac{u}{2} D_aY^a\right)\,,\\
     \xi_\text{BMS}^a &= Y^{a} -\frac{1}{r} D^a \left(T+\dfrac{u}{2}D_b Y^b\right)\,.
\end{align}
\end{subequations}
The coordinate transformation generated by the above vector fields form the symmetries of the space of solutions which are parametrized by a time-independent function $T(\theta^a)$ generating super-translations, and a time-independent vector $Y^a(\theta^b)$ on the sphere generating super-Lorentz transformations. These form the celebrated generalized BMS algebra~\cite{Campiglia:2014yka, Campiglia:2015yka,  Compere:2018ylh, Campiglia:2020qvc, Campiglia:2016hvg} (\textit{i.e.}, the smooth version of~\cite{deBoer:2003vf, Barnich:2009se, Barnich:2010eb, Barnich:2011mi}). The modification of the metric under the BMS group reads\footnote{We have $\Delta D_aY^a = D_a(\Delta Y^a - Y^a )$. Note that the Ricci tensor $R_{ab}=\gamma_{ab}$ on the unit sphere.}
\begin{subequations} \label{BMSmetric}
\begin{align} \label{BMSmetricuu}
    \delta_\text{BMS}\, g_{uu} &=-G\bigl(\Delta+2\bigr)\dot{f} + \mathcal{O}(G^2)=- \dfrac{G}{2}\,D_a\Bigl( Y^a + \Delta Y^a\Bigr)+ \mathcal{O}(G^2)\,,\\
     \delta_\text{BMS}\, g_{ua} &=- \dfrac{G}{2}D_a\bigl(\Delta+2\bigr) f + \mathcal{O}(G^2) \nonumber\\ &=-G D_a\left[ T + \frac{1}{2}\Delta T + \frac{u}{4} D_b\Bigl( Y^b + \Delta Y^b\Bigr)\right]+ \mathcal{O}(G^2),\\
     \delta_\text{BMS}\, g_{ab} &=  G \Bigl[2r^2\, D_{\langle a} Y_{b\rangle} - 2r\, D_{\langle a} D_{b\rangle}f \Bigr]+ \mathcal{O}(G^2)\,, \label{BMSmetricab}
\end{align}
\end{subequations}
where we recall that $D_{\langle a} Y_{b\rangle}=D_{(a} Y_{b)}-\frac{1}{2}\gamma_{ab}D_c Y^c$ and $D_{\langle a} D_{b\rangle} f = D_{a} D_{b} f-\frac{1}{2}\gamma_{ab}\Delta f$. The transformation law of the asymptotic metric on the sphere $q_{ab}$ defined from $g_{ab}= r^{2}q_{ab}+\mathcal{O}(r^1)$ agrees with Eq.~(2.20) of~\cite{Compere:2018ylh}. We note that the leading $uu$ component of the metric is given by Eq.~\eqref{BMSmetricuu} where the divergence $D_a Y^a$ only involves the determinant of the metric on the unit  sphere. This is consistent with Eqs.~(2.5) and (2.25) or Eqs.~(3.11) and (3.21) of~\cite{Compere:2018ylh}. 

It is worth pointing out that the kernel of the operator $\Delta+2$ appearing in the BMS transformation of the $uu$ component of the metric in Eq.~\eqref{BMSmetricuu} is the $\ell=1$ harmonics, \textit{i.e.} $(\Delta+2)f=0$ if and only if $f$ is made of the $\ell=1$ harmonics. Similarly the kernel of the operator $D_{\langle a} D_{b\rangle}$ appearing in the BMS transformation of the $ab$ component~\eqref{BMSmetricab} [see also the shear~\eqref{BMSshear}] is the $\ell=0$ and $\ell=1$ harmonics. In order to make this explicit, we decompose the function $f$ into STF spherical harmonics
\begin{equation}
f = T+\frac{u}{2}D_aY^a = \sum_{\ell=0}^{+\infty} n_L f_L(u)\,,
\end{equation}
where the STF coefficients $f_L$ are linear functions of $u$, and find
\begin{subequations}
	\begin{align}
	\bigl(\Delta +2\bigr)f &= - \sum_{\ell=0}^{+\infty} (\ell+2)(\ell-1)\,n_L f_L\,,\\
	D_{\langle a} D_{b\rangle}f &= e_{\langle a}^i e_{b\rangle}^j \sum_{\ell=0}^{+\infty} \ell(\ell-1) \,n_{L-2} f_{ijL-2}\,.\label{DaDb}
	\end{align}
\end{subequations}

For completeness, we can now detail the boundary conditions at spatial infinity that could be imposed in order to completely fix the asymptotic frame, even though we will not enforce these conditions in the following sections since they remove the generalized BMS asymptotic symmetry group at spatial infinity \cite{Strominger:2013jfa,Troessaert:2017jcm,Henneaux:2018cst,Compere:2020lrt}. First, upon fixing the boundary metric to be the unit sphere metric, $q_{ab}=\gamma_{ab}$, all proper super-Lorentz transformations are discarded and the generalized BMS algebra reduces to the original BMS algebra. Second, upon imposing stationarity in the asymptotic past $u \rightarrow -\infty$, one sets the momenta to zero, $P_i=0$ and the boost are discarded. Since the Bondi news $N_{ab}$ is zero or decays in the asymptotic past, the electric part of the Bondi shear defined as $C^+$ in the decomposition $C_{ab}=-2GD_{\langle a}D_{b\rangle }C^+ +2G\eps_{c (a}D_{b)}D^cC^-$ satisfies $\lim_{u\to-\infty}C^+=C(\theta,\phi)$ with $C \rightarrow C + T$ under a supertranslation. One can then discard all supertranslations but the Poincar\'e translations by fixing all harmonics $\ell > 1$ of $C$. On the other hand, the rotations not aligned with the total angular momentum can be discarded by setting the Bondi angular momentum $N_a(u=-\infty)$ to canonical form, $N_a=-3 J \sin^2\theta \p_a \phi$. Finally, the spatial translations can be discarded by setting the mass dipole to zero, $M_i=0$, which is equivalent to choosing the center-of-mass frame. The BMS symmetry group is then gauge-fixed  to $\mathbb R \times SO(2)$, the symmetry group of asymptotically stationary solutions consisting of time translations and rotations around the axis of the total angular momentum. In conclusion, one can reduce the four-dimensional diffeomorphism group to $\mathbb R \times SO(2)$ after imposing Newman-Unti gauge~\eqref{eq:NU gauge0}, boundary gauge fixing conditions~\eqref{eq:bndgauge} and additional boundary conditions at spatial infinity as just described.

\subsection{Bondi data to linear order}\label{sec:aspects}

Finally, we shall write the metric~\eqref{eq:metric} including the bulk terms in the form 
\begin{subequations}\label{linearmetric}
\begin{align}
g_{uu} &= -1 - G\bigl(\Delta +2\bigr)\dot f + \frac{2G}{r}\biggl[ m + \sum_{k=1}^{+\infty} \frac{1}{r^k} K_k\biggr] + \mathcal{O}(G^2)\,,\label{linmetric-uu}\\
g_{ua} &= G\Biggl(\frac{1}{2} D_b C_{a}^{b} + \frac{1}{r}\biggl[\frac{2}{3}N_a + e_a^i\sum_{k=1}^{+\infty} \frac{1}{r^k} P^i_k\biggr]\Biggr) + \mathcal{O}(G^2)\,,\label{linmetric-ua}\\
g_{ab} &=  r^2 \Biggl[ \gamma_{ab} + 2GD_{\langle a} Y_{b\rangle} + \frac{G}{ r} \left( C_{ab} + e_{\langle a}^i e_{b\rangle}^j \sum_{k=1}^{+\infty} \frac{1}{r^k} Q^{ij}_k\right)\Biggr] + \mathcal{O}(G^2)\,.\label{linmetric-ab}
\end{align}
\end{subequations}
The sub-dominant contributions in $1/r$ in the metric~\eqref{linearmetric} read as
\begin{subequations}\label{subdominant}
	\begin{align}
	K_k &= \frac{1}{(k+1)(k+2)} \sum_{\ell=k}^{+\infty}\frac{(\ell+1)(\ell+2)}{\ell!} \,a_{k\ell} \, n_L M_L^{(\ell-k)} + \mathcal{O}(G)\,,\\
	P^i_k &= \frac{2}{k+3}\sum_{\ell=k+1}^{+\infty}\frac{\ell+2}{\ell!} \,a_{k+1\ell} \, n_{L-1} \Bigl[M_{iL-1}^{(\ell-k-1)}+\frac{2\ell}{\ell+1}\varepsilon_{ipq}n_p S_{qL-1}^{(\ell-k-1)}\Bigr] + \mathcal{O}(G)\,,\\
	Q^{ij}_k &= 4 \frac{k-1}{k+1} \sum_{\ell=k}^{+\infty}\frac{1}{\ell!} \,a_{k\ell} \,n_{L-2} \Bigl[M_{ijL-2}^{(\ell-k)}+\frac{2\ell}{\ell+1}\varepsilon_{ipq}n_p S_{jqL-2}^{(\ell-k)}\Bigr] + \mathcal{O}(G)\,.
	\end{align}
\end{subequations}
Note that $Q^{ij}_1=0$ for $k=1$. Therefore, at linear order in $G$ the next order correction term in $1/r$ in the metric $ g_{ab}$ beyond the shear $C_{ab}$ is absent. This is just a feature of the linear metric, since at quadratic order $\mathcal{O}(G^2)$ there is a well-known term quadratic in the shear. 

To leading order when $r\to \infty$ the metric \eqref{linearmetric} is defined by the so-called Bondi mass aspect $m$, angular momentum aspect $N_a$ and shear $C_{ab}$ (see \textit{e.g.}~\cite{Tamburino:1966zz, Bonga:2018gzr,Compere:2019gft}). These are functions of time $u$ and the angles $\th^a$. The mass and angular momentum aspects are given in terms of the multipole moments to linear order in $G$ by
\begin{subequations}\label{eq:aspects}
	\begin{align}
	m &= \sum_{\ell=0}^{+\infty} \dfrac{(\ell+1)(\ell+2)}{2\ell!} \,n_L M_{L}^{(\ell)} + \mathcal{O}(G)\,,\label{eq:maspect}\\
	N_{a} &= e_{a}^i \sum_{\ell=1}^{+\infty}\frac{(\ell+1)(\ell+2)}{2(\ell-1)!} \,n_{L-1}\left[ M_{iL-1}^{(\ell-1)}+\frac{2\ell}{\ell+1}\varepsilon_{ipq}n_p S_{qL-1}^{(\ell-1)}\right] + \mathcal{O}(G)\,.\label{eq:AMaspect}
	\end{align}
\end{subequations}

In the next section we will work out the mass loss and angular momentum loss formulas for $\dot m$ and $\dot N_a$ to quadratic order in $G$. But we already note that
\begin{equation}\label{Ndot}
\dot{N}_{a} = D_a m + e_{a}^i \sum_{\ell=1}^{+\infty}\frac{\ell(\ell+2)}{(\ell-1)!} \,\varepsilon_{ipq} n_{pL-1} S_{qL-1}^{(\ell)} + \mathcal{O}(G)\,,
\end{equation}
in agreement with the Einstein equation for the angular momentum aspect.

To define the shear we introduce the usual asymptotic waveform in transverse-trace-free (TT) gauge, given in terms of the multipole moments by (see \textit{e.g.}~\cite{Blanchet:2013haa})
\begin{equation}\label{HTT}
H^{ij}_\text{TT} = 4 \!\perp_\text{TT}^{ijkl}\sum_{\ell=2}^{+\infty} \frac{n_{L-2}}{\ell!}\left[ M_{klL-2}^{(\ell)}-\frac{2\ell}{\ell+1}\varepsilon_{kpq}n_p S_{lqL-2}^{(\ell)}\right] + \mathcal{O}(G)\,,
\end{equation}
where $\perp_\text{TT}^{ijkl}$ is the TT projection operator. Then the shear is given by
\begin{equation}\label{BMSshear}
C_{ab} = e_{\langle a}^i e_{b\rangle}^j H^{ij}_\text{TT} - 2D_{\langle a}D_{b\rangle} f \,.
\end{equation}
The first term is directly related to the usual two polarization waveforms at infinity. Posing $H_+=\lim_{r\to\infty}(r h_+)$ and $H_\times=\lim_{r\to\infty}(r h_\times)$ we have\footnote{We adopt for the polarization vectors $\varepsilon_\th^i=e_\th^i$ and $\varepsilon_\varphi^i=e_\varphi^i/\sin\th$ such that $\varepsilon_\th^i \varepsilon_\th^j + \varepsilon_\varphi^i \varepsilon_\varphi^j = \perp^{ij} = \gamma^{ab}e_a^i e_b^j$.} 
\begin{equation}
e_{\langle a}^i e_{b\rangle}^j H_\text{TT}^{ij} = \begin{pmatrix} H_+ & H_\times \sin\th \\ H_\times \sin\th & - H_+ \sin^2\th \end{pmatrix} \,.
\end{equation}
The second term in Eq.~\eqref{BMSshear} comes from the BMS transformation as
\begin{equation}\label{transfBMSshear}
\delta_\text{BMS}\, C_{ab} = -2\,D_{\langle a} D_{b\rangle}f\,.
\end{equation}

In the stationary limit, the Bondi mass and angular momentum aspects reduce to the conserved ADM mass $M$ and angular momentum $N_a=3e_a^{i}\epsilon_{ipq}n_pS_q$, and the shear $C_{ab}$ vanishes up to the supertranslation shift~\eqref{transfBMSshear} with $f=T$. In the metric~\eqref{linearmetric}, the canonical multipole moments $M_L$, $S_L$ appear in $g_{uu}$, $r^{-1}g_{ua}$, $r^{-2}g_{ab}$ exactly at order $r^{-\ell+1}$ and match (up to a normalisation) with the standard Geroch-Hansen multipole moments~\cite{Geroch:1970cd,Hansen:1974zz,Thorne:1980ru,1983GReGr..15..737G}. In the zero supertranslation frame (\textit{i.e.} $\Delta(\Delta+2)T=0$) and in a Lorentz frame (\textit{i.e.} $D_{\langle a}Y_{b \rangle }=0$), the stationary limit of Eq.~\eqref{linearmetric} is, modulo $\mathcal{O}(G^2)$,
\begin{subequations}\label{linearmetricstat}
\begin{align}
g_{uu}^\text{stat} &= -1 + 2G \sum_{\ell=0}^{+\infty} \frac{(-)^\ell}{\ell!} M_L \partial_L\!\left(\frac{1}{r}\right) \,,\\
g_{ua}^\text{stat} &= -2 G e_a^i\sum_{\ell=1}^{+\infty} \frac{(-)^\ell}{\ell!} (2\ell-1) \Bigl[M_{iL-1} + \frac{2\ell}{\ell+1}\varepsilon_{ipq}n_p S_{qL-1}\Bigr]\partial_{L-1}\!\left(\frac{1}{r}\right) \,,\\
g_{ab}^\text{stat} &=  r^2 \gamma_{ab} + 4 G e_{\langle a}^i e_{b\rangle}^j \sum_{\ell=2}^{+\infty} \frac{(-)^\ell}{\ell!}\frac{(\ell-1)(2\ell-1)(2\ell-3)}{\ell+1}\times\nn\\
&\qquad\qquad\qquad\qquad\qquad\times\Bigl[M_{ijL-2} + \frac{2\ell}{\ell+1}\varepsilon_{ipq}n_p S_{jqL-2}\Bigr]\partial_{L-2}\!\left(\frac{1}{r}\right) \,.
\end{align}
\end{subequations}

\section{Newman-Unti metric to quadratic order}
\label{sec:quadratic}

At second order in $G$ the perturbation~\eqref{eq:PM} reads as\footnote{It implies $	\sqrt{\vert \tl g\vert} = 1+\tfrac{G}{2}h_1 +G^2 \Bigl(  \tfrac{1}{2}h_2 +\tfrac{1}{8}h_1^2 -\frac{1}{4}h_1^{\rho\sigma}h_{1 \rho\sigma} \Bigr)+ \mathcal{O}(G^3)$ and\vspace{-10pt} \begin{align*}
\tl g_{\mu\nu} &=\eta_{\mu\nu} \!+\!G\left(-h_{1 \mu\nu}+\tfrac{1}{2}h_1 \eta_{\mu\nu} \right) \!+\!G^2\Bigl[- h_{2 \mu\nu}-\tfrac{1}{2}h_1 h_{1 \mu\nu} + \left(\tfrac{1}{2}h_2 + \tfrac{1}{8}h_1^2 -\tfrac{1}{4}h_1^{\rho\sigma}h_{1 \rho\sigma}\right)\eta_{\mu\nu}+ h_{1 \mu \rho}h^{\;\rho}_{1\;\nu}\Bigr] \!+\!\mathcal{O}(G^3)\,, \\ 
\tl g^{\mu\nu} &=\eta^{\mu\nu} + G\left(h_1^{ \mu\nu}-\frac{1}{2}h_1 \eta^{\mu\nu} \right) + G^2\left[h_2^{\mu\nu}-\frac{1}{2}h_2 \eta^{\mu\nu} -\frac{1}{2}h_1 h_1^{ \mu\nu} + \left(\frac{1}{8}h_1^2 +\frac{1}{4}h_1^{\alpha\beta}h_{1 \alpha\beta}\right)\eta^{\mu\nu} \right]+\mathcal{O}(G^3).
\end{align*}}
\begin{equation}
    \sqrt{\vert \tl g\vert} \tilde g^{\mu\nu} = \eta^{\mu\nu} + h^{\mu\nu} = \eta^{\mu\nu} + G h_1^{\mu\nu} + G^2 h_2^{\mu\nu} + \mathcal{O}(G^3). 
\end{equation}
In the following we will denote $h_{2} \equiv  \eta_{\mu\nu}h^{\mu\nu}_{2}$ and the indices are lowered by the background Minkowski metric $\eta_\mn$. At second order in $G$, the NU gauge conditions \eqref{eq:NU gauge} imply the following equations for the functions $U_2$, $R_2$, $\Theta^a_2$, respectively,
\begin{subequations}\label{eq:quadorder}\vspace{-10pt}
\begin{align}
\tl k^{\mu}\tl \partial_\mu U_2 &= \frac{1}{2}\tl k_{\mu}\tl k_{\nu}h_2^{\mu\nu} + \left(\frac{1}{2}\tl\partial^{\mu}U_1 - \tl k_\nu h_1^{\mu\nu}\right)\tl\partial_{\mu}U_1\,,\label{eq:eqU2}\\
\tl k^{\mu}\tl \partial_\mu R_2 &= \frac{1}{8}h_1^2 -\frac{1}{4} h_1^{\mu\nu}h_{1 \mu\nu} + \frac{1}{2} h_2 + \tl n_i\left[ \tl \p_i U_2 -\tl k_\m  h_{2}^{\mu i} + (\tl \p_\m U_1)h_1^{\m i} \right]\nonumber\\ & \quad+ \left(\tl\partial^{\mu}U_1 - \tl k_\nu h_1^{\mu\nu}\right)\tl\partial_{\mu}R_1\,,\label{eq:eqR2}\\
\tl k^{\mu}\tl \partial_\mu \Theta^a_2 &= \frac{\tl e^a_i}{\tl r}\left[\tl \p_i U_2 - \tl k_\mu h_2^{\mu i}+(\tl \p_\m U_1)h_1^{\mu i} \right] + \left(\tl\partial^{\mu}U_1 - \tl k_\nu h_1^{\mu\nu}\right)\tl\partial_{\mu}\Theta^a_1\,.\label{eq:eqTheta2}
\end{align}
\end{subequations}
See Appendix~\ref{app:PM} for a formal generalization of these equations to any PM order.

In the following, we will show how the explicit solution for the quadratic metric in harmonic coordinates, \textit{i.e.}, solving the Einstein field equations~\eqref{eq:EFE} to order $G^2$ for some given multipole interactions, can be used as an input in our algorithm in order to generate the corresponding Bondi-NU metric.

The main features of the quadratic metric in harmonic coordinates are~\cite{Blanchet:1992br, Blanchet:1997ji, Blanchet:1997jj}: (i) the presence of gravitational-wave tails, corresponding to quadratic interactions between the constant mass $M$ and varying multipole moments $M_L$ and $S_L$ (for $\ell\geqslant 2$); (ii) the mass and angular momentum losses describing the corrections of the constant ADM quantities introduced in the linear metric ($M$ and $S_i$) due to the GW emission;\footnote{Similarly there are corrections associated with the losses of linear momentum (or recoil) and the position of the center of mass, see \textit{e.g.}~\cite{Blanchet:2018yqa,Compere:2019gft}.} (iii) the presence of the non-linear memory effect. We investigate the effects (i) and (ii) in the subsections below but postpone (iii) to future work. 

\subsection{Tails and the mass-quadrupole interaction}\label{sec:tail}

In this subsection we construct the NU metric corresponding to the monopole-quadrupole interaction $M\times M_{ij}$, starting from the explicit solution in harmonic coordinates given by (see Appendix B of~\cite{Blanchet:1992br}, or Eq.~(2.8) of~\cite{Blanchet:1997jj}) 
\begin{subequations}\label{eq:MMij}
	\begin{align}
	h^{00}_2 &= M \tl n_{pq} \,\tl r^{-4} \left\{ -21 M_{pq}
	-21 \tl r M^{(1)}_{pq} + 7 \tl r^2 M^{(2)}_{pq}
	+ 10 \tl r^3 M^{(3)}_{pq} \right\} \nonumber\\
	&+ 8 M \tl n_{pq} \int^{+\infty}_1 \dd x \,Q_2 (x) M^{(4)}_{pq}(\tl t -  \tl r x)\,,\\
	%%%%%%%%%%%%%%%%%%%%%%%%%%%%%%%%%%%%%%%%%%%%%%%%%%%%%%%%%%%%%%%
	h^{0i}_2 &= M \tl n_{ipq} \,\tl r^{-3} \left\{
	-M^{(1)}_{pq}  - \tl r M^{(2)}_{pq} - {1\over 3} \tl r^2 M^{(3)}_{pq} \right\}  \nonumber\\
	&+ M \tl n_p \,\tl r^{-3} \left\{ -5 M^{(1)}_{pi} - 5 \tl r M^{(2)}_{pi} + {19\over 3} \tl r^2
	M^{(3)}_{pi} \right\} \nonumber\\
	&+ 8 M \tl n_p \int^{+\infty}_1 \dd x \,Q_1 (x) M^{(4)}_{pi} (\tl t - \tl r x) \,,\\
	%%%%%%%%%%%%%%%%%%%%%%%%%%%%%%%%%%%%%%%%%%%%%%%%%%%%%%%%%%%%
	h^{ij}_2 &= M \tl n_{ijpq} \,\tl r^{-4} \left\{ -{15\over 2}
	M_{pq} - {15\over 2} \tl r M^{(1)}_{pq} - 3 \tl r^2 M^{(2)}_{pq} - {1\over 2} \tl r^3
	M^{(3)}_{pq} \right\} \nonumber\\
	&+ M \delta_{ij} \tl n_{pq} \,\tl r^{-4} \left\{ -{1\over 2} M_{pq}
	- {1\over 2} \tl r M^{(1)}_{pq} - 2 \tl r^2
	M^{(2)}_{pq} - {11\over 6} \tl r^3 M^{(3)}_{pq}
	\right\} \nonumber\\
	&+ M \tl n_{p(i} \,\tl r^{-4} \left\{ 6 M_{j)p} + 6 \tl r M^{(1)}_{j)p}
	+ 6 \tl r^2 M^{(2)}_{j)p} + 4 \tl r^3 M^{(3)}_{j)p} \right\} \nonumber\\
	&+ M \,\tl r^{-4} \left\{ - M_{ij} - \tl r M^{(1)}_{ij} - 4
	\tl r^2 M^{(2)}_{ij} - {11\over 3} \tl r^3 M^{(3)}_{ij} \right\} \nonumber\\
	&+ 8 M \int^{+\infty}_1 \dd x \,Q_0 (x) M^{(4)}_{ij} (\tl t - \tl r x) \,. 
	\end{align}
\end{subequations}
The metric is composed of two types of terms: the so-called ``instantaneous'' ones depending on the quadrupole moment $M_{ij}$ and its derivatives at time $\tl u =\tl t-\tl r$, and the ``hereditary'' tail terms depending on all times from $- \infty$ in the past until $\tl u$. The tail integrals are expressed in Eq.~\eqref{eq:MMij} by means of the Legendre function of the second kind $Q_\ell$ (with branch cut from $-\infty$ to $1$), given by the explicit formula in terms of the Legendre polynomial $P_\ell$: 
\begin{equation}\label{eq:Qell}
Q_{\ell} (x) = {1\over 2} P_\ell (x) \ln
\left(\frac{x+1}{x-1} \right)- \sum^{ \ell}_{ j=1}
\frac{1}{j} P_{\ell -j}(x) P_{j-1}(x) \,.
\end{equation}
We recall that the Legendre function $Q_\ell$ behaves like $1/x^{\ell+1}$ when $x \to +\infty$, and that its leading expansion when $y\equiv x-1\to 0^+$ reads (with $H_\ell=\sum^\ell_{j=1}{1\over j}$ being the $\ell$th harmonic number)
\begin{equation}\label{eq:limQell}
Q_\ell(1+y) = -{1\over 2} \ln \left( {y\over 2}\right) - H_\ell + \mathcal{O}\left(y \ln y\right) \,.
\end{equation}

With the known harmonic metric~\eqref{eq:MMij} [or see below Eq.~\eqref{eq:MML}], we apply our algorithm to generate the Bondi-NU metric. We focus on the case of the mass-quadrupole interaction $M\times M_{ij}$, keeping track of all instantaneous and tail terms. Plugging $h^{\mu\nu}_2$ given by Eq.~\eqref{eq:MMij} as well as $h^{\mu\nu}_1$ and $U_1$ given in the previous section in the right-side of Eq.~\eqref{eq:eqU2}, and retaining only the mass-quadrupole interaction we get 
\begin{align}\label{eq:eqU2expl}
\tl k^{\mu}\tl \partial_\mu U_2 &= M \tl n_{pq} \Bigl[-6 \tl r^{-4} M_{pq} - 3 \tl r^{-3} M^{(1)}_{pq} + 6 \tl r^{-2} M^{(2)}_{pq} \Bigr] \nonumber\\
&+ 4 M \tl n_{pq} \int^{+\infty}_1 \dd x \left[Q_2-2Q_1+Q_0-\frac{1}{2}\right] M^{(4)}_{pq}(\tl t - \tl rx)\,. 
\end{align} 
We remark that, as an intermediate step to obtain~\eqref{eq:eqU2expl}, an instantaneous term of the form $-2\tl r^{-1}M \tilde n_{pq}M^{(3)}_{pq}$ has been equivalently written as the last term in the second line. In this form, it is explicit that the integrand of Eq.~\eqref{eq:eqU2expl} does not diverge in the limit $x\to 1^+$, despite the logarithmic pole, thanks to the factor $(x-1)$ in the sum of Legendre functions,
\begin{equation}
Q_2(x)-2Q_1(x)+Q_0(x)-\frac{1}{2} = \frac{1}{4}(x-1)\left[(3x-1)\,\ln
\left(\frac{x+1}{x-1} \right) - 6\right]\,. 
\end{equation}
This permits to immediately integrate Eq.~\eqref{eq:eqU2expl} over $\tl r$ (while keeping $\tl u$ fixed) with result\footnote{Note that $\tl t - \tl rx = \tl u - \tl r(x-1)$ and $\partial_{\tl r}M^{(3)}[\tl u - \tl r (x-1)]\big\vert_{\tl u=\text{const}}=-(x-1)M^{(4)}[\tl u - \tl r (x-1)]$.}
\begin{align}\label{eq:solU2}
U_2 &= M \tl n_{pq} \biggl[2 \tl r^{-3} M_{pq} + \frac{3}{2} \tl r^{-2} M^{(1)}_{pq} -  \int^{+\infty}_1 \dd x ~(3x-1)\ln
\left(\frac{x+1}{x-1} \right) M^{(3)}_{pq}(\tl t - \tl rx)\biggr] \,. 
\end{align} 
In principle this is valid up to an homogeneous solution corresponding to a linear gauge transformation starting to order $G^2$. It will be of the form $-\xi^u_2=-f_2$ where $f_2$ is a function of $\tl u=\tl t-\tl r$ and $\tl \theta^a$. It thus takes the same form as the linear gauge transformation already introduced to order $G$ in Eq.~\eqref{eq:U1}. Hence, we can absorb $f_2$ into the redefinition of $f$ through the replacement $f\rightarrow f+G f_2$, and the solution~\eqref{eq:solU2} is the most general in our setup. Following the same procedure outlined above to compute $U_2$, we obtain
\begin{subequations}
\begin{align} \label{eq:solR2}
R_2 &= M \tl n_{pq} \biggl[ \tl r^{-2} M^{(1)}_{pq} + \frac{9}{2} \tl r^{-1} M^{(2)}_{pq}  - 3  \int^{+\infty}_1 \dd x \ln
\left(\frac{x+1}{x-1} \right) M^{(3)}_{pq}(\tl t - \tl rx)\biggr]\,,\\\Theta_2^a \!&=\! M \frac{\tl n_p \tl e^a_q}{\tl r} \bigg[ \tl r^{-3} M_{pq}+\frac{2}{3}\tl r^{-2}M^{(1)}_{pq} +2\tl r^{-1} M^{(2)}_{pq} +2  \int_{1}^{+\infty}\! \! \!\!\dd  x ~(x-1)\ln\left(\frac{x+1}{x-1}\right) \! M^{(3)}_{pq}(\tl t - \tl rx)\bigg].
\end{align}
\end{subequations}

Having determined $U_2$, $R_2$ and $\Theta_2^a$ we continue our algorithm and successively obtain the contravariant components $g^{rr}$, $g^{ra}$ and $g^{ab}$ of the NU metric, and then its covariant components $g_{uu}$, $g_{ua}$ and $g_{ab}$, see Sec.~\ref{sec:algo}. We consistently keep only the terms corresponding to the mass-quadrupole $M \times M_{ij}$ interaction. In the end we recall that we have to express the metric components in terms of the NU coordinates $x^\mu=(u,r,\theta^a)$ by applying (the inverse of the) coordinate transformation~\eqref{eq:transfBondi}. In order to present the result in the best way we introduce the following tail-modified quadrupole moment as defined by~\cite{Blanchet:2013haa}\footnote{We have changed the integration variable to $z = r (x-1)$. In previous formul\ae, it is convenient to decompose $\text{ln}(\frac{x+1}{x-1})=-\text{ln}(\frac{z}{2\mathcal{P}})+\text{ln}(1+\frac{z}{2r})+\text{ln}(\frac{r}{\mathcal{P}})$, where $\mathcal{P}$ is the constant introduced in Eq.~\eqref{eq:U1R1Th1}. The first term gives the tail in the quadrupole~\eqref{eq:radquad}, the second term gives the tail in the metric~\eqref{eq:guu2}--\eqref{eq:guagab2} and the third term is cancelled after reexpressing the metric in NU coordinates.} 
\begin{align}\label{eq:radquad}
{M}^\text{rad}_{ij}(u)={M}_{ij}(u)+2 G M \int_{0}^{+\infty} \dd z \left[\ln \left(\frac{z}{2\mathcal{P} }\right)+\frac{11}{12}\right]{M}_{ij}^{(2)}(u-z) + \mathcal{O}\left(G^2\right)\,.
\end{align}
Such definition agrees with the expression of the radiative quadrupole moment parametrizing the leading $r^{-1}$ piece of the metric at future null infinity. Restoring the powers of $c^{-1}$ we see that the tail provides a 1.5PN correction $\sim c^{-3}$ to the quadrupole. Generally the radiative quadrupole moment is rather defined as the second-time derivative of ${M}^\text{rad}_{ij}$, see Eq.~(76a) of~\cite{Blanchet:2013haa}. But here, as we not only control the leading term $r^{-1}$ but also all the subleading terms $r^{-2}$, $r^{-3}$, \textit{etc.} in the expansion of the metric at infinity, it will turn out to be better to define the radiative moment simply as ${M}^\text{rad}_{ij}$. 

We find that the NU metric $g_{uu}$ to quadratic order $G^2$ for the mass-quadrupole interaction, including all terms in the expansion at infinity, reads 
\begin{align}\label{eq:guu2}
g_{uu} &= -1 + G \left[2 M r^{-1} + 6 r^{-1} n^{ij}\Mijrad{2} + 6 r^{-2} n^{ij}\Mijrad{1} + 3 r^{-3} n^{ij} M^\text{rad}_{ij}\right] \\
&\qquad\quad +\frac{3}{2} G^2 M r^{-3} n^{ij}\biggl[ ~\Mijrad{1} + r^{-2}\int_{0}^{+\infty} \!\dd z \, \frac{M^\text{rad}_{ij}(u-z)}{\left(1+\frac{z}{2r}\right)^2}\biggr] +\mathcal{O}\left(G^3\right)\,.\nonumber
\end{align}
We recover Eq.~\eqref{eq:metricuu} for the linear part, and we see that to quadratic order the tails nicely enter the metric only through the replacement of the canonical moment $M_{ij}$ by the radiative moment $M^\text{rad}_{ij}$ defined by Eq.~\eqref{eq:radquad}. In fact, with this approximation (neglecting $G^3$ terms), we can use either $M_{ij}$ or $M^\text{rad}_{ij}$ in the second line of Eq.~\eqref{eq:guu2}. 

Note that the last term of Eq.~\eqref{eq:guu2}, involving a time integral over the radiative moment, is ``exact'' all over the exterior region of the source. The integral is convergent under our assumption of stationarity in the past. Furthermore, this term is of order $\mathcal{O}(r^{-4})$ at null infinity where it admits an expansion involving only powers of $r^{-1}$. We have the regular expansion when $r\to+\infty$ for $u=\text{const}$:	
\begin{align}\label{eq:regular}
&\int_{0}^{+\infty} \!\dd z \, \frac{M^\text{rad}_{ij}(u-z)}{(1+\frac{z}{2r})^2} = \sum_{p=0}^{+\infty} \frac{(-)^p (p+1)}{(2r)^{p}} \int_0^{u+\mathcal{T}} \!\dd z \,z^{p} M^\text{rad}_{ij}(u-z)+\frac{4r^2 M^\text{rad}_{ij}(-\mathcal{T})}{2r+u+\mathcal{T}}\,\\
&\qquad\qquad\qquad = \sum_{p=0}^{+\infty} \frac{(-)^p (p+1)}{(2r)^{p}} \int_0^{+\infty}\!\dd z \,z^{p} \Bigl[M^\text{rad}_{ij}(u-z) - M^\text{rad}_{ij}(-\mathcal{T})\Bigr]+ 2 r M^\text{rad}_{ij}(-\mathcal{T})\,,\nonumber
\end{align}
where $-\mathcal{T}$ is the finite instant in the remote past before which the multipoles are constant.

Further processing we obtain the other components of the NU metric as 
\begin{subequations}\label{eq:guagab2}
	\begin{align}
	g_{ua} &= G e_a^i n^j\biggl\{ -2\,\Mijrad{2} + 2 r^{-1}\Big( \varepsilon_{ijk}S_k+2\Mijrad{1}\Big) + 3 r^{-2} M^\text{rad}_{ij} \\
	&\qquad\quad + \frac{1}{2} G M\biggl[ 3r^{-2}\Mijrad{1} + r^{-4} \int_{0}^{+\infty} \dd z \,\frac{5 + \frac{3z}{2r}}{\left(1 + \frac{z}{2r}\right)^3} \,M^\text{rad}_{ij}(u-z)\biggr]\biggr\} +\mathcal{O}\left(G^3\right)\,,\nonumber\\
	%%%%%%%%%%%%%%%%%%%%%%%%%%%%%%%%%%%%%%%%%%%%%%%%%%%%%%%%%%%%%
	g_{ab} &= r^2\Biggl[ \gamma_{ab} + 2G e_{\langle a}^i e_{b\rangle}^j\left( r^{-1}\Mijrad{2} + r^{-3} M^\text{rad}_{ij}\right) \\
	&\quad + G^2 M e_{\langle a}^i e_{b\rangle}^j \biggl( r^{-3} \Mijrad{1} + \frac{1}{4} r^{-5} \int_{0}^{+\infty} \dd z \,\frac{18 + \frac{8z}{r} + \frac{z^2}{r^2}}{\left(1 + \frac{z}{2r}\right)^4}\,{M}^\text{rad}_{ij}(u-z)\biggr) \Biggr] +\mathcal{O}\left(G^3\right)\,.\nonumber
	\end{align}
\end{subequations}
Again we find some remaining tail integrals, but which rapidly fall off when $r\to\infty$ and admit an expansion in simple powers of $r^{-1}$. Finally we conclude that the expansion of the NU metric at infinity is regular, without the powers of $\ln r$ which plague the expansion of the metric in harmonic coordinates. In intermediate steps of the computation, however, logarithmic divergences occur in the quadratic term, but they are cancelled by the expansion of the linear term taking into account $\tilde u = u + 2GM \ln (r/\mathcal{P}) + \mathcal{O}(G^2)$.

The fact that the NU metric admits a regular (smooth) expansion when $r \to +\infty$ to all orders, without logarithms, is nicely consistent with the earlier work~\cite{Blanchet:1986dk} which proved the property of asymptotic simplicity in the sense of Geroch and Horowitz~\cite{GH78}, \textit{i.e.}, with a smooth conformal boundary at null infinity, for the large class of radiative coordinate systems, containing the Bondi and NU coordinates. Indeed, a crucial assumption in the proof of~\cite{Blanchet:1986dk} as well as in our work, see Eq.~\eqref{eq:regular}, is that the metric is stationary in the past (for $u\leqslant-\mathcal{T}$).

To second order in $G$, as already commented, we could still add to the construction some arbitrary homogeneous solutions of the equations for $U_2$, $R_2$ and $\Theta^a_2$, but the corresponding terms in the metric will have exactly the same form as those found to linear order in $G$, see Eqs.~\eqref{eq:deltametric}, and shown to describe with appropriate boundary conditions the modification of the metric under the BMS group. 

From the results~\eqref{eq:guu2}--\eqref{eq:guagab2}, one can easily deduce the mass and angular momentum aspects $m$ and $N_a$, and the Bondi shear $C_{ab}$, for the case of the mass-quadrupole interaction to order $G^2$. As expected the Bondi data are entirely determined by the radiative quadrupole moment~\eqref{eq:radquad}. Recalling the expression of the metric in the NU gauge as given by Eqs.~\eqref{eq:mCabdef} and~\eqref{eq:Nadef} in Appendix~\ref{app:map}, where $N_a$ is defined according to the convention of~\cite{Flanagan:2015pxa}, we get 
\begin{subequations}\label{eq:mNC2}
\begin{align}
m &= M + 3n^{ij} \,\Mijrad{2} + \mathcal{O}\left(G^2\right)\,,\label{eq:m2}\\
N_a &= 3 e_a^i n^j\Bigl(\varepsilon_{ijk}S_k+2 \Mijrad{1}\Bigr) + \mathcal{O}\left(G^2\right)\,,\\
C_{ab} &= 2 e_{\langle a}^i e_{b\rangle}^j \,\Mijrad{2} + \mathcal{O}\left(G^2\right)\,.
\end{align}
\end{subequations}
We have added in the angular momentum aspect the linear contribution due to the total constant (ADM) angular momentum or spin $S_i$, as read off from Eq.~\eqref{eq:AMaspect}. 

Notice that the difference between the Newman-Bondi and Bondi radii is a term quadratic in $C_{ab}$, see Eq.~\eqref{eq:diffNU-B}. This term is thus quadratic in the source moment $M_{ij}$, and so, for the mass-quadrupole interaction $M\times M_{ij}$ considered in this section, there is no difference between the NU and Bondi gauges.

In the stationary limit, the Bondi mass and angular momentum aspects as well as the shear~\eqref{eq:mNC2} reduce to their linear expressions. Moreover, the radiative quadrupole $M_{ij}^{\text{rad}}$ as defined in Eq.~\eqref{eq:radquad} reduces to the canonical one $M_{ij}$. More generally, it follows from dimensional analysis that no perturbative non-linear correction exists to the Bondi data or to the multipole moments in the stationary case. Indeed, suppose a non-linear correction to the moment $M_L$, built from $n$ moments $M_{L_1}$, $\cdots$,  $M_{L_n}$. In the stationary case this correction must be of the type $\sim \frac{G^{n-1}}{c^{2n-2}} M_{L_1}\cdots M_{L_n}$ with $\ell = n-1 + \sum \ell_i$ in order to match the dimension. Furthermore, we must also have $\sum \ell_i = \ell + 2k$ for the correspondence of indices, where $k$ is the number of contractions among the indices $L_1\cdots L_n$. The two conditions are clearly incompatible. This entails that the canonical multipoles $M_L,S_L$ agree with the Geroch-Hansen multipoles \cite{Geroch:1970cd,Hansen:1974zz} at the non-linear level.

We can in principle generalize the latter results to multipole interactions $M \times M_L$ and $M \times S_L$ (with any $\ell\geqslant 2$), starting from the known expressions of tail terms in the metric in harmonic coordinates:\footnote{This is a straightforward generalization of the mass quadrupole tail terms in Eq.~\eqref{eq:MMij}.}
\begin{subequations}\label{eq:MML}
	\begin{align}
	&h^{00}_2 = 16 M \,\frac{\tl n_{L}}{\ell!} \int^{+\infty}_1 \dd x \,Q_{\ell} (x) \,M^{(\ell+2)}_{L}(\tl t - \tl r x) + \cdots\,,\\
	%%%%%%%%%%%%%%%%%%%%%%%%%%%%%%%%%%%%%%%%%%%%%%%%%%%%%%%%%%%%%%%
	&h^{0i}_2 =  16 M \,\frac{\tl n_{L-1}}{\ell!} \int^{+\infty}_1 \dd x \left[ Q_{\ell-1}(x) \,M^{(\ell+2)}_{iL-1} - \frac{\ell}{\ell+1} \,Q_{\ell}(x) \,\varepsilon_{ipq} \,\tl n_p \,S^{(\ell+2)}_{qL-1}\right] + \cdots\,,\\
	%%%%%%%%%%%%%%%%%%%%%%%%%%%%%%%%%%%%%%%%%%%%%%%%%%%%%%%%%%%%
	&h^{ij}_2 = 16 M \,\frac{\tl n_{L-2}}{\ell!} \int^{+\infty}_1 \dd x \left[ Q_{\ell-2}(x) \,M^{(\ell+2)}_{ijL-2} - \frac{2\ell}{\ell+1} \,Q_{\ell-1}(x) \,\tl n_p \,\varepsilon_{pq(i}S^{(\ell+2)}_{j)qL-2}\right] + \cdots\,.
	\end{align}
\end{subequations}
Here the ellipsis refer to many non-tail contributions, in the form of instantaneous (\textit{i.e.}, local-in-time) terms depending on the multipole moments only at time $\tl u$. Considering the previous results we can conjecture that the mass and angular momentum aspects will take the same form as in Eqs.~\eqref{eq:aspects} but with the canonical moments $M_L$ and $S_L$ replaced by the radiative moments $M_L^\text{rad}$ and $S_L^\text{rad}$~\cite{Blanchet:1995fr}
\begin{subequations}\label{eq:radtailL}
\begin{align}
{M}^\text{rad}_{L}(u) &= {M}_{L}(u)+2 G M \int_{0}^{+\infty} \dd z \left[\ln \left(\frac{z}{2\mathcal{P} }\right)+\kappa_\ell\right]{M}_{L}^{(2)}(u-z) + \mathcal{O}\left(G^2\right)\,,\\
{S}^\text{rad}_{L}(u) &= {S}_{L}(u)+2 G M \int_{0}^{+\infty} \dd z \left[\ln \left(\frac{z}{2\mathcal{P} }\right)+\pi_\ell\right]{S}_{L}^{(2)}(u-z) + \mathcal{O}\left(G^2\right)\,,
\end{align}
\end{subequations}
where the constants are given by (with $H_\ell=\sum^\ell_{j=1}{1\over j}$) 
\begin{equation}
	\kappa_\ell = \frac{2\ell^2+5\ell+4}{\ell(\ell+1)(\ell+2)}+H_{\ell-2}\,,\qquad\pi_\ell = \frac{\ell-1}{\ell(\ell+1)}+H_{\ell-1}\,.
\end{equation}
More work would be needed to generalize our algorithm in order to include any multipole interactions $M \times M_L$ and $M \times S_L$ (especially instantaneous ones).

\subsection{Mass and angular momentum losses}\label{sec:loss}

Taking the angular average of the mass aspect $m$ we obtain the Bondi mass $M_\text{B}\equiv\int\frac{\dd\Omega}{4\pi} \,m$. At this stage, we find from Eqs.~\eqref{eq:m2} or~\eqref{eq:maspect} that the Bondi mass just equals the ADM mass $M_\text{ADM}\equiv M$. This is because we have not yet included the mass loss by GW emission which arises in this formalism from the quadratic interaction between two quadrupole moments, say $M_{ij}\times M_{kl}$, as well as higher multipole moment interactions. The losses of mass and angular momentum are straightforward to include in the formalism, starting from the known results in harmonic coordinates.

The terms responsible for mass and angular momentum losses (at the lowest quadrupole-quadrupole interaction level) in the harmonic-coordinate metric are (see \textit{e.g.} Eq.~(4.12) in~\cite{Blanchet:1997ji}):
\begin{subequations}\label{eq:lossharmcoord}
\begin{align}
h_2^{00} &= \frac{4}{5\tl r} \int_{-\infty}^{\tl u} \!\dd v \,M_{pq}^{(3)}M_{pq}^{(3)}(v) + \cdots \,,\\
h_2^{0j} &= \frac{4}{5} \varepsilon_{jpq}\tl\partial_p\left(\frac{1}{\tl r}\varepsilon_{qrs}\int_{-\infty}^{\tl u} \!\dd v \,M_{rt}^{(2)} M_{st}^{(3)}(v) \right) + \cdots \,,\\
h_2^{jk} &= \cdots\,,
\end{align}
\end{subequations}
where again, the ellipsis denote many instantaneous (local-in-time) terms, in contrast with the non-local time anti-derivative integrals over the multipole moments in Eq.~\eqref{eq:lossharmcoord}. Importantly, the ellipsis in Eq.~\eqref{eq:lossharmcoord} also contain another type of non-local terms that are associated with the non-linear memory effect, but which we shall not discuss  here. The complete quadrupole-quadrupole interaction $M_{ij}\times M_{kl}$ has been computed in harmonic coordinates in~\cite{Blanchet:1997ji}, including the description of the various GW losses and the non-linear memory effect.

We thus apply our algorithm to generate the corresponding mass and angular momentum losses in the NU metric. In this calculation we only keep track of the non-local-in-time (or ``hereditary'') integrals, and neglect all the instantaneous terms. Furthermore, as we said we do not consider the memory effect, which is disconnected from GW losses (see \textit{e.g.}~\cite{Blanchet:1997ji}). Finally we are restricted to the quadrupole-quadrupole interaction, as in Eq.~\eqref{eq:lossharmcoord}. 

Looking at the second-order equations~\eqref{eq:quadorder} we see that we are just required to solve
\begin{subequations}\label{eq:tosolve}
\begin{align}
\tl k^{\mu}\tl \partial_\mu U_2 &= \frac{1}{2}\tl k_{\mu}\tl k_{\nu}h_2^{\mu\nu} + \cdots\,,\\
\tl k^{\mu}\tl \partial_\mu R_2 &= \frac{1}{2} h_2 + \tl n_i\left[ \tl \p_i U_2 -\tl k_\m  h_{2}^{\mu i} \right] + \cdots\,,\\
\tl k^{\mu}\tl \partial_\mu \Theta^a_2 &= \frac{\tl e^a_i}{\tl r}\left[\tl \p_i U_2 - \tl k_\mu h_2^{\mu i} \right] + \cdots\,.
\end{align}
\end{subequations}
We obtain successively (changing consistently harmonic to NU coordinates)
\begin{subequations}\label{eq:URT2}
	\begin{align}
U_2 &= \frac{2}{5}\ln(r/\mathcal{P}) \int_{-\infty}^{u} \!\dd v \,M_{pq}^{(3)}M_{pq}^{(3)}(v) + \cdots\,,\\
R_2 &= \cdots\,,\\
\Theta^a_2 &= - \frac{2}{5} \frac{e^a_i n^j}{r^2} \,\varepsilon_{ijp}\,\varepsilon_{pqr} \int_{-\infty}^{u} \!\dd v \,M_{qs}^{(3)}M_{rs}^{(2)}(v)+ \cdots\,.
	\end{align}
\end{subequations}
We find no such hereditary terms in $R_2$. The logarithmic term in $U_2$ corrects the light cone deviation at linear order as given by Eq.~\eqref{eq:U1}. The corresponding contributions in the NU metric follow as
\begin{subequations}\label{eq:NUlosses}
	\begin{align}
	g_{uu} &= -1 - \frac{2G}{5} r^{-1} \int_{-\infty}^{u} \!\dd v \,M_{pq}^{(3)}M_{pq}^{(3)}(v) + \cdots\,,\\
	g_{ua} &= - \frac{4G}{5} \,\frac{e^a_i n^j}{r} \,\varepsilon_{ijp}\,\varepsilon_{pqr} \int_{-\infty}^{u} \!\dd v \,M_{qs}^{(2)}M_{rs}^{(3)}(v)+ \cdots\,,\\
	g_{ab} &= r^2 \gamma_{ab}\left[1 - \frac{2G}{5} r^{-1} \int_{-\infty}^{u} \!\dd v \,M_{pq}^{(3)}M_{pq}^{(3)}(v)\right] + \cdots\,.
\end{align}
\end{subequations}
Combining this with previous results~\eqref{eq:m2} or~\eqref{eq:maspect} we obtain the mass aspect which is now accurate enough to include the physical GW mass loss
\begin{equation}
m = M + 3n^{ij} \,\Mijrad{2} - \frac{G}{5} \int_{-\infty}^{u} \!\dd v \,M_{pq}^{(3)}M_{pq}^{(3)}(v) + \cdots\,.
\end{equation}
Hence the Bondi mass $M_\text{B}=\int\frac{\dd\Omega}{4\pi} \,m$ reads (where $M$ is the constant ADM mass)
\begin{equation}\label{eq:massB}
M_\text{B} = M - \frac{G}{5} \int_{-\infty}^{u} \!\dd v \,M_{pq}^{(3)}M_{pq}^{(3)}(v) + \cdots\,.
\end{equation}
The mass loss in the right-side is characterized by the hereditary (or ``semi-hereditary'')\footnote{We distinguish~\cite{Blanchet:1992br} semi-hereditary integrals that are just time anti-derivatives of products of multipole moments as in Eq.~\eqref{eq:massB}, from truly hereditary integrals extending over the past,  like the tail terms in Eq.~\eqref{eq:MML}.} non-local integral, in contrast with the instantaneous contributions indicated by dots. Such instantaneous terms will be in the form of total time derivatives in the corresponding flux balance equation, and may be neglected in average over a typical orbital period for quasi-periodic systems. Thus the averaged balance equation reduces to 
\begin{equation}\label{eq:balance}
\langle \frac{\dd M_\text{B}}{\dd t} \rangle = - \frac{G}{5} \,M_{pq}^{(3)} M_{pq}^{(3)}\,,
\end{equation}
which is of course nothing but (with this approximation) the balance equation corresponding to the standard Einstein quadrupole formula.

In a similar way we obtain the angular momentum aspect and Bondi shear as
\begin{subequations}\label{eq:amB}
\begin{align}
N_a &= 6 e_a^i n^j \left[ \frac{1}{2} \varepsilon_{ijp} S_p + \Mijrad{1} - \frac{G}{5} \varepsilon_{ijp} \varepsilon_{pqr}\int_{-\infty}^{u} \!\dd v \,M_{qs}^{(2)}M_{rs}^{(3)}(v) + \cdots\right]\,,\label{eq:N2}\\
C_{ab} &= 2 e_{\langle a}^i e_{b\rangle}^j \,\Mijrad{2} - \frac{2G}{5}\gamma_{ab} \int_{-\infty}^{u} \!\dd v \,M_{pq}^{(3)}M_{pq}^{(3)}(v) + \cdots\,.
\end{align}
\end{subequations}
The Bondi angular momentum is defined from the angular momentum aspect by 
\begin{equation}
S_i^\text{B} \equiv \frac{1}{2}\varepsilon_{ipq}\int\frac{\dd\Omega}{4\pi} \,e_a^p \,n^q \Bigl(N_a-\frac{\alpha}{4G}C_{ab}D_c C^{bc}\Bigr)\,.
\end{equation}
As shown in~\cite{Compere:2019gft}, this quantity requires a prescription for $\alpha$ which is fixed to $\alpha=1$ in~\cite{Barnich:2011mi, Flanagan:2015pxa, Distler:2018rwu, Compere:2018ylh}, $\alpha=0$ in~\cite{Pasterski:2015tva, Hawking:2016sgy} or $\alpha=3$ in~\cite{Bonga:2018gzr}. Since the $\alpha$-term  gives instantaneous terms as well as higher order terms, we can simply ignore it for this computation. Hence we have
\begin{equation}\label{eq:SBondi}
S^\text{B}_i = S_i - \frac{2G}{5} \varepsilon_{ipq} \int_{-\infty}^{u} \!\dd v \,M_{ps}^{(2)}M_{qs}^{(3)}(v) + \cdots\,.
\end{equation}
Upon averaging this leads to the usual quadrupole balance equation for angular momentum\footnote{The angular momentum aspect itself satisfies, see also Eq.~\eqref{Ndot},
$$\frac{\dd N_{a}}{\dd t} = D_a m + 3 e_{a}^i\,\varepsilon_{ipq} n_{p} \frac{\dd S_{q}^\text{B}}{\dd t} + \cdots\,.$$}
\begin{equation}\label{eq:balanceS2}
\langle \frac{\dd S^\text{B}_i}{\dd t}\rangle = - \frac{2G}{5} \varepsilon_{ipq}  M_{ps}^{(2)}M_{qs}^{(3)} \,.
\end{equation}
Note that the discussion of the GW losses in the linear momentum (or recoil) and the center-of-mass position would require the coupling between the mass quadrupole and the mass octupole moments, which is outside the scope of the present calculation.

\section{Conclusion and perspectives}
\label{sec:concl}

In this paper we have shown how to implement practically the transformation of the metric of an isolated matter source in the MPM (multipolar post-Minkowskian) approach from harmonic (de Donder) coordinates to Bondi-like NU (Newman-Unti) coordinates. This is of interest because the asymptotic properties of radiative space-times are generally discussed within the Bondi-Sachs-Penrose formalism, while the connection to the source's properties is done by a matching procedure to the source using the MPM expansion. 

In particular we obtain explicit expressions for the NU metric valid at any order in the radial distance to the source (while staying outside the domain of the source), expressed in terms of the canonical mass and current multipole moments. Under the assumption of stationarity in the remote past, we prove that the NU metric (for particular multipole moment couplings) admits a regular expansion at future null infinity. This is consistent with the fact that the MPM expansion satisfies the property of asymptotic simplicity~\cite{Blanchet:1986dk}.

On the other hand the canonical moments are known in terms of the source's parameters to high PN (post-Newtonian) order. Our approach permits to rewrite explicit results derived in harmonic coordinates using the MPM approximation into the Bondi-Sachs-Penrose formalism for the asymptotic structure, including the notions of Bondi shear, and mass and angular momentum aspects. In particular, we recover from our construction the generalized BMS (Bondi-van der Burg-Metzner-Sachs) residual symmetry group leaving invariant the NU metric under appropriate boundary conditions at future null infinity.\footnote{By contrast, harmonic coordinates are preserved by a distinct residual symmetry group which includes the Poincar\'e group as well as multipole symmetries whose associated Noether charges are the canonical multipole moments~\cite{Compere:2017wrj}.}

To non-linear order our construction is in principle valid for any coupling between the canonical moments. In this paper we have worked out the coupling between the mass and the quadrupole, including the contributions due to non-local (hereditary) tail effects but also all local (instantaneous) terms. Including the non-local (semi-hereditary) terms arising from the coupling between two quadrupoles, we obtain the mass and angular momentum losses due to the GW emission through the expressions of the mass and angular momentum aspects. However we ignored all the instantaneous terms in the quadrupole-quadrupole metric, as well as the contributions from the non-linear memory effect. In future work we intend to thoroughly investigate the quadrupole-quadrupole interaction in our framework, and in particular discuss the occurrence of the non-linear memory effect, thereby contrasting the perspective from approximation methods in harmonic coordinates with that from asymptotic studies in Bondi-like coordinates confined close to future null infinity. 

\paragraph{Acknowledgments}
R.O. and A.S. are grateful to Bernard Whiting for enlightening discussions on related topics. G.C. acknowledges Y. Herfray and A. Puhm for interesting discussions. G.F., R.O. and A.S. would like  to thank the  Munich Institute for Astro- and Particle Physics  (MIAPP), which is  funded by the Deutsche Forschung-sgemeinschaft  (DFG, German Research Foundation)  under Germany's Excellence Strategy --  EXC-2094 -- 390783311, for giving them the opportunity of  preliminary  discussions  that triggered the  current project. R.O. and A.S. thank the Institut d'Astrophysique de Paris for the hospitality when this work was initiated and the COST Action GWverse CA16104 for partial financial support. R.O. is funded by the European Structural and Investment Funds (ESIF) and the Czech Ministry of Education, Youth and Sports (MSMT), Project CoGraDS - CZ.02.1.01/0.0/0.0/15003/0000437. A.S. receives funding from the European Union's Horizon 2020 research and innovation program under the Marie Sklodowska-Curie grant agreement No 801505. G.C. is Senior Research Associate from the Fonds de la Recherche Scientifique F.R.S.-FNRS (Belgium) and he acknowledges support from the FNRS research credit J.0036.20F, bilateral Czech convention PINT-Bilat-M/PGY R.M005.19 and the IISN convention 4.4503.15. 

\appendix

%----------------------------------------------------------------------------
\section{Map between Bondi and Newman-Unti gauges}
\label{app:map}

Bondi gauge and Newman-Unti gauge differ by a choice of radial coordinate~\cite{Barnich:2011ty}. They both admit identical asymptotic symmetry groups, phase spaces and physical quantities~\cite{Barnich:2011ty}. We denote in both coordinate systems the angular coordinates as $\theta^a$ and the coordinate labelling the foliation of null hypersurfaces as $u$. Let us refer to $r_\text{B}$ as the Bondi radius and $r_\text{NU}$ as the Newman-Unti radius. The Newman-Unti radius $r_\text{NU}$ is the affine parameter along the outgoing null rays, while the Bondi radius is the luminosity distance such that $\p_{r_\text{B}} [\text{det}(g_{ab})/r_\text{B}^4]=0$. There are certain advantages of NU coordinates over the Bondi coordinates, in particular the bulk extension of NU is larger than Bondi~\cite{Madler:2016xju}. The relationship between the radii is given by~\cite{Barnich:2011ty}
\begin{equation}
r_\text{NU}=r_\text{B} + \int_{r_\text{B}}^\infty dr' \bigl(g_{r_\text{B} u}+1\bigr),\qquad r_\text{B} = \left( \frac{\text{det}\, g_{ab}}{\text{det}\, \gamma_{ab}}\right)^{1/4}\,.
\end{equation}
For large radii, we have 
\begin{subequations}\label{eq:diffNU-B}
\begin{align}
r_\text{NU} &= r_\text{B} + \frac{1}{16 r_\text{B}} C_{ab}C^{ab}+\mathcal{O}(r_\text{B}^{-2})\,,\\
r_{B} &= r_\text{NU} - \frac{1}{16 r_\text{NU}} C_{ab}C^{ab}+\mathcal{O}(r_\text{NU}^{-2})\,. 
\end{align}
\end{subequations}
The deviation only starts from order $1/r_\text{B}$ or $1/r_\text{NU}$. We deduce that $C_{ab}$ and $m$ can be read off from the metric in Newman-Unti gauge as 
\begin{subequations}\label{eq:mCabdef}
\begin{align}
g_{uu}^{\text{NU}} &= -1 + \frac{2m_\text{NU}}{r_\text{NU}}+ \mathcal{O}(r_\text{NU}^{-2})\,, \\
g_{ab}^{\text{NU}} &= r_\text{NU}^2 \gamma_{ab}+r_\text{NU} C_{ab} + \mathcal{O}(r_\text{NU}^0)\,,
\end{align}
\end{subequations}
with $m_\text{NU}=m+\frac{1}{16}\partial_u (C_{ab}C^{ab})$. Instead, 
\begin{equation}\label{eq:change}
g_{ua}^{\text{NU}} =  g_{ua}^{\text{B}} + \frac{1}{16 r} D_a ( C_{bc}C^{bc})+ \mathcal{O}(r^{-2})\,,
\end{equation}
where $r$ is either $r_\text{B}$ or $r_\text{NU}$. In the convention of~\cite{Flanagan:2015pxa}, the angular momentum aspect $N_a$ is read in Bondi gauge from 
\begin{equation}
g^{\text{B}}_{ua} = \frac{1}{2}D^b C_{ab}+\frac{1}{r}\left[\frac{2}{3} N_a - \frac{1}{16 }D_a \left(C_{bc}C^{bc}\right) \right] +\mathcal{O}(r^{-2})\,.
\end{equation}
We deduce from Eq.~\eqref{eq:change} that it is read in Newman-Unti gauge from 
\begin{equation}\label{eq:Nadef}
g^{\text{NU}}_{ua} = \frac{1}{2}D^b C_{ab}+\frac{2}{3r} N_a   +\mathcal{O}(r^{-2}) \,.
\end{equation}

\section{Equations for any PM order}
\label{app:PM}

At any given PM order $p\in \mathbb{N}$, the NU gauge conditions \eqref{eq:NU gauge} imply the following equations for $U_p$, $R_p$ and $\Theta^a_p$, respectively,
\begin{subequations}
	\begin{align}
	\tl k^{\m}\tl \p_\m U_p &= \frac{1}{2}\tl k_\m \tl k_\n h_p^{\m\n} + \sum_{\substack{m,n \geqslant 1 \\ m+n=p}}\Bigl(\frac{1}{2}\tl\partial^{\mu}U_m - \tl k_\nu h_m^{\mu\nu}\Bigr)\tl\partial_{\mu}U_n + \frac{1}{2}\sum_{\substack{m,n,q \geqslant 1 \\ m+n+q=p}}(\tl\partial_{\nu}U_m)(\tl\partial_{\mu}U_n)h_q^{\m\n}\,,\\
	%%%%%%%%%%%%%%%%%%%%%%%%%%%%%%%%%%%%%%%%%%%%%%%%%%%%%%%%
	\tl k^{\m}\tl \p_\m R_p &= \sum_{\substack{m\geqslant 1 \\ m+n=p}} \binom{\frac{1}{2}}{m}\biggl[\sum_{n\geq 1}\vert \tl g\vert_n\biggr]^m + \tl n_i\biggl[ \tl \p_i U_p -\tl k_\m  h_{p}^{\mu i} + \sum_{\substack{m,n \geqslant 1 \\ m+n=p}}(\tl \p_\m U_n)h_m^{\m i} \biggr] + \nonumber\\
	\quad& + \sum_{\substack{m,n \geqslant 1 \\ m+n=p}}\left(\tl\partial^{\mu}U_m - \tl k_\nu h_m^{\mu\nu}\right)\tl\partial_{\mu}R_n  + \sum_{\substack{m,n,q \geqslant 1 \\ m+n+q=p}}(\tl\partial_{\nu}U_m)(\tl\partial_{\mu}R_n)h_q^{\m\n}\,,\\
	%%%%%%%%%%%%%%%%%%%%%%%%%%%%%%%%%%%%%%%%%%%%%%%%%%%%%%%
	\tl k^{\m}\tl \p_\m \Theta^a_p &= \frac{\tl e^a_i}{\tl r}\biggl[\tl \p_i U_p - \tl k_\mu h_p^{\mu i}+ \sum_{\substack{m,n \geqslant 1 \\ m+n=p}}(\tl \p_\m U_n)h_m^{\mu i} \biggr] + \sum_{\substack{m,n \geqslant 1 \\ m+n=p}}\left(\tl\partial^{\mu}U_m - \tl k_\nu h_m^{\mu\nu}\right)\tl\partial_{\mu}\Theta^a_n + \nonumber\\
	&\quad + \sum_{\substack{m,n,q \geqslant 1 \\ m+n+q=p}}(\tl\partial_{\nu}U_m)(\tl\partial_{\mu}\Theta^a_n)h_q^{\m\n}\,.
	\end{align}
\end{subequations}

To derive the equation for $R_p$, one formally writes
\begin{equation}
\vert \tl g\vert = 1 + \sum_{n\geqslant 1} G^n \vert \tl g\vert_n ~\longrightarrow~ \sqrt{\vert \tl g\vert} = \sum_{m\geqslant 0} \binom{\frac{1}{2}}{m}\biggl[\sum_{n\geqslant 1}G^{n}(\vert \tl g\vert)_n\biggr]^m\,.
\end{equation}

\bibliography{ref_multipoles}
%--------------------------------------------------------------------------------------------------
%--------------------------------------------------------------------------------------------------
\end{document}